\documentclass[10pt]{article}
\usepackage{geometry}
\usepackage{cancel,amssymb,amsmath,mathrsfs,upgreek}
\usepackage{ textcomp }
\usepackage{ulem}
\usepackage{mathdots}
\usepackage{color,tikz}

\usetikzlibrary{patterns, decorations.markings,arrows, calc}
\usepackage{color,todonotes}
\usepackage{framed}
\usepackage[colorlinks=true, pdfstartview=FitV, linkcolor=darkblue, citecolor=darkblue, urlcolor=darkblue]{hyperref}
\definecolor{shadecolor}{rgb}{0.98, 0.98, 0.9}
\definecolor{darkgreen}{rgb}{0.2, 0.5,  0}
\definecolor{darkblue}{rgb}{0.1,0.1,0.45}
\definecolor{red}{rgb}{0.9,0,0}

\def \bea#1\eea {\begin{align} #1 \end{align}}

\def\&{\hspace{-15pt}&}
\def\nn{\nonumber}
\def \scr{\mathscr}
\def \u{\mathfrak A}
\def \CCC{{\varkappa}}

\def \K{ \mathscr K}
\def\ds{\displaystyle}
\def \P {\mathbb P}
\def\ov{\overline}
\def \mb {\mathbb }
\def\bt{\mathbf t}
\def\bfs{\mathbf s}
\def \eqref#1{(\ref{#1})}

\def \remove #1 { \sout{ {\color{red} #1}}}

\def \wt{\widetilde}
\def\Cauchy{ \mathbf {C}}
\def \wh{\widehat}
\newcommand{\C}{\mathbb{C}}
\renewcommand{\le}{\left}
\newcommand{\ri}{\right}
\newcommand{\R}{\mathbb{R}}
\newcommand{\Z}{\mathbb{Z}}

\newcommand{\1}{\mathbf{1}}
\newcommand{\g}{\gamma}

\renewcommand{\d}{\mathrm d}
\renewcommand{\mod}{\,\mathrm{mod}\,}
\newcommand{\pa}{\partial}

\def\res{\mathop{\mathrm {res}}\limits_}

\def\be{\begin{equation}}
\def\ee{\end{equation}}

\def\bg{\begin{gathered}}
\def\eg{\end{gathered}}

\newtheorem{theorem}{Theorem}[section]

\newtheorem{example}[theorem]{Example}

\newtheorem{lemma}[theorem]{Lemma}
\newtheorem{remark}[theorem]{Remark}
\newtheorem{problem}[theorem]{Riemann--Hilbert Problem}

\newtheorem{proposition}[theorem]{Proposition} 
\newtheorem{corollary}[theorem]{Corollary} 
 
\newtheorem{definition}[theorem]{Definition}

\def\bet
{
\begin{shaded}
\begin{theorem}
}
\def\eet{\end{theorem} \end{shaded}}

\def\bp
{
\begin{shaded}
\begin{proposition}
}
\def\ep{
\end{proposition}\end{shaded}
}
\def\QED {\hfill $\blacksquare$\par\vskip 10pt}

\def\CC{\mathcal{C}}

\def \DD{\mathbb D}

\def\g{\gamma}

\renewcommand{\theequation}{\arabic{section}.\arabic{equation}}

\makeatletter
\@addtoreset{equation}{section}
\makeatother

\begin{document}
\vspace{0.2cm}
\begin{center}
\begin{Large}
\textbf{Pad\'e\ approximants on Riemann surfaces and KP tau functions} 
\end{Large}
\end{center}

\begin{center}
M. Bertola$^{\dagger\ddagger\diamondsuit}$ \footnote{Marco.Bertola@\{concordia.ca, sissa.it\}}, 
\\
\bigskip
\begin{minipage}{0.7\textwidth}
\begin{small}
\begin{enumerate}
\item [${\dagger}$] {\it  Department of Mathematics and
Statistics, Concordia University\\ 1455 de Maisonneuve W., Montr\'eal, Qu\'ebec,
Canada H3G 1M8} 
\item[${\ddagger}$] {\it SISSA, International School for Advanced Studies, via Bonomea 265, Trieste, Italy }
\item[${\diamondsuit}$] {\it Centre de recherches math\'ematiques,
Universit\'e de Montr\'eal\\ C.~P.~6128, succ. centre ville, Montr\'eal,
Qu\'ebec, Canada H3C 3J7}
\end{enumerate}
\end{small}
\end{minipage}
\vspace{0.5cm}
\end{center}

\begin{abstract}
The paper has two relatively distinct  but connected goals; the first is to define the notion of Pad\'e\ approximation of Weyl-Stiltjes transforms  on an arbitrary compact  Riemann surface of higher genus. The data consists of a contour in the Riemann surface and a measure on it,  together with the additional datum of a local coordinate near a point and a divisor of degree $g$. 
The denominators of the resulting Pad\'e--like approximation also satisfy an orthogonality relation and are sections of appropriate line bundles. 
A Riemann--Hilbert problem for a square matrix of rank two is shown to characterize these orthogonal sections, in a similar fashion to the ordinary orthogonal polynomial case.

The second part extends this idea to explore its connection to integrable systems. The same data can be used to define a pairing between two sequences  of line bundles. The locus in the deformation space where the pairing becomes degenerate for fixed degree coincides with the zeros of a ``tau'' function. We show how this tau function satisfies the Kadomtsev--Petviashvili hierarchy with respect to either deformation parameters, and a certain modification of the 2--Toda hierarchy when considering the whole sequence of tau functions.
We also show how this construction is related to the Krichever construction of algebro--geometric
solutions. 
\end{abstract}

\setcounter{tocdepth}{2}
\normalem
\tableofcontents

\section{Introduction}
The theory of Hermite--Pad\'e\ approximation is  intimately connected with the theory of (mutliple) orthogonal polynomials. The prototypical of these connections is as follows: one considers a measure $\d\mu$ with finite moments on the real axis and its Stiltjes transform 
\be
\label{Weyl0}
W(z) = \int_\R \frac {\d\mu(x)}{z-x}.
\ee
Then we find polynomials $Q_{n-1}, P_{n}$ (of the degree suggested by the subscript) such that $W(z) = \frac {Q_{n-1}(z)}{P_n(z)} + \mathcal O(z^{-2n-1})$ as $|z|\to \infty$ (in the sense of asymptotic expansion). A simple computation shows that the denominators $P_n$ are orthogonal polynomials in the sense that 
\be
\nn
\int_{\R} \d\mu(x) P_n(x) P_m(x) = 0\ \ \ n\neq m.
\ee
While this connection is classical, a more recent result \cite{FIK2,FIK} connects the construction of the orthogonal polynomials with a Riemann--Hilbert problem. This connection was instrumental in the theory of random matrices to provide the first rigorous proof of several universality results \cite{DKMVZ}.

If the measure is made to depend on (formal) parameters $\d\mu(x;{\bf t}) = {\rm e}^{\sum_{j\geq 0} t_j x^j} \d\mu(x;0) $ then the recurrence coefficients of the polynomials $P_n(x; {\bf t})$ provide a solution to the Toda lattice equations (see for example the review in \cite{Deift}). Furthermore, the Hankel determinant of the corresponding moments 
\be
\nn
\Delta_n({\bf t}) = \det\bigg[\int_\R x^{a+b-2} \d\mu(x; {\bf t})\bigg]_{a,b=1}^n
\ee 
provide tau functions for the Kadomtsev--Petviashvili hierarchy.
This type of interplay between (multiple)  Pad\'e\ approximation (and related multiple orthogonality)
and integrable systems has been exploited in numerous papers, to name a few \cite{ls-cubicstring, Bertola:CauchyMM,  Bertola:MomentTau, Gekhtman:Elementary,Kuijlaars:RHPBOPs,AdlerVanMoerbeke}.

On a seemingly disconnected track, the theory of integrable systems, notably the theory of the Kadomtsev--Petviashvili  (KP) hierarchy is famously intertwined with the theory of Riemann surfaces \cite{Krichever, SegalWilson} in the class of algebro-geometric solutions. 

There seems to be little or no literature attempting to connect the worlds of Pad\'e\ approximation and the algebro-geometric setup. 

The present paper is a first foray in the sparsely populated landscape between these areas.

On the side of Pad\'e\ approximation theory, we mention the recent work \cite{olver} where the authors consider a sequence of functions on elliptic curves with antiholomorphic involution  which are orthogonal with respect to a measure on the fixed ovals. The setup is comparable to, but not the same as, the class of examples we consider here in Section \ref{antihol}.  We could not find other literature which is relevant to our present approach.

Before describing the results we add a few words of caution: by the nature of this paper there are potentially two classes of mathematicians that could be interested. On one side the community of approximation theory and on the other side the community of integrable systems. Inevitably here we are obliged to use certain notions of the theory of Riemann surfaces that are rather common in the integrable-system community but less so in the approximation theory one. The author is leaning more towards the first and therefore the language used in the paper tends to reflect this bias. 
I have tried to clarify certain terminology wherever possible.

\paragraph{Description of results.}
We fix a Riemann surface $\CC$ of genus $g\geq 1$, and a divisor of degree $g$ (i.e. a collection of points counted with multiplicity so that the total number  is $g$). On $\CC$ we fix a contour $\gamma$ and a weight differential $\d\mu(q)$ (details are in Sec. \ref{secWeyl}).
We choose a distinguished point on $\CC$ which we denote by $\infty$ (since it plays the role of the point at infinity in the complex plane). The last piece of data is a choice of local coordinate $z:\mathcal U\setminus \{\infty\}\to \C$ in a neighbourhood $\mathcal U$ of $\infty$ such that $\lim_{p\to\infty} z(p)=\infty$.
The main results are listed below:
\begin{itemize}
\item 
We start from the description of a suitable extension of the Pad\'e\ approximation problem for Weyl-Stiltjes transforms in higher genus. The Weyl-Stiltjes function, $W$, analog to \eqref{Weyl0}, is defined in terms the given data in Def. \ref{DefWeyl}; the definition requires the use of a suitable Cauchy kernel which replaces the expression ${\frac {1}{z-x}}$. In fact the object we define is not a ``function'' but a  holomorphic differential on $\CC\setminus \gamma\cup \{\infty\}$ with a jump discontinuity along $\g$ equal to  the chosen measure. Further motivation for this choice is descriped in Sec. \ref{secWeyl}.
\item We define the Pad\'e\ approximation problem in Def. \ref{defPade}:
instead of a ratio of polynomials the relevant generalization requires the ratio of a meromorphic differential $\mathfrak Q_{n-1}$ and a meromorphic function $P_n$ such that it approximates the Weyl-Stiltjes function at the point $\infty\in\CC$ to appropriate order.  The denominators $P_n$ are shown to be orthogonal (in the sense of non-Hermitean orthogonality) with respect to the measure $\d\mu$ on $\gamma$. 
\item One of the most versatile tools for the study of asymptotic of orthogonal polynomials has proven to be the formulation in terms of a Riemann--Hilbert problem (RHP) \cite{FIK,FIK2, Deift}. For this reason we formulate the precise analog in this context in Sec. \ref{secRHP}. The situation for higher genus curves is, expectedly, more complicated: the RHP is still a problem for a  $2\times 2$ matrix but the existence of the solution is not sufficient to guarantee its uniqueness (contrary to what happens in genus zero). The existence and uniqueness of the solution is, however, equivalent to the non-vanishing of a determinant $D_n$ \eqref{nothankel} which generalizes the Hankel determinant of the moments: this is Theorem \ref{thmDn}. 
\item The familiar determinantal expression and Heine formulas for orthogonal polynomials have a strict counterpart in \eqref{pn} and Prop. \ref{propheine}, respectively. 
\item In Section \ref{KPsect} we consider a generalization of the relation between biorthogonal polynomials and KP tau functions/random matrices \cite{AdlerVanMoerbeke}. With the choice of a local coordinate $1/z(p)$ near the point $\infty\in \CC$ we have the same data (curve, line bundle, local coordinate) which was used by Krichever to construct algebro--geometric solutions.
We define two sequences of biorthogonal sections of certain line bundles and a pairing between them in terms of integration along the curve $\gamma$ with the given measure.
The tau function is defined in Def. \ref{defTau}: it depends on an integer $n$ (the dimension of the spaces of sections being paired) and for $n=0$ it factorizes into the product of two algebro--geometric KP tau functions. For general $n>0$ the tau function vanishes only if the pairing is degenerate or the Krichever line bundle is special. This tau function depends on two infinite sets of ``times''. We prove (Thm. \ref{mainKP}) that it is a KP tau function (Def. \ref{defHBI}) in both sets of times. The proof uses the general Hirota bilinear relations in integral formulation. We compute explicitly the Baker and dual-Baker functions in terms of the bi-orthogonal sections of the two line bundles (Prop. \ref{propBaker}, \ref{propdualBaker}). We finally identify the sequence as an instance of a (suitably modified) solution of the $2$--Toda hierarchy, as presented in \cite{AdlerVanMoerbeke, UenoTakasaki}.
\end{itemize}
%
%
%\begin{enumerate}
%\item KP + Riemann surfaces
%\item KP + OPs
%\item KP + random matrices
%\item OPs and RHP (FIK)
%\item OPs on Riemann surfaces: only Olver?
%\end{enumerate}
%
%Suppose that $\tau(\bt)$ is a tau function for the Kadomtsev--Petviashvili hierarchy (a KP tau function for short) \cite{SegalWilson}; then it is a well known fact that 
%\be
%\tau\le(\bt + \sum_{j\leq n} [x_j^{-1}]\ri) \exp\le[\sum_{j\leq n} \sum_{\ell\geq 0} t_\ell x_j^\ell\ri] =\frac{ \det \le[\pa_{t_1}^{a-1} \Psi(x_b;\bt)\ri]_{a,b=1}^{n}}{\prod_{a<b} (x_a-x_b)}
%\ee

\section{Weyl-Stiltjes function of a measure}
\label{secWeyl}
Let $\CC$ be a Riemann surface of genus $g$ and $\scr D$ a non-special divisor on $\CC$ of degree $g$. Let $\infty\in \CC\setminus \scr D$. 
%\blue
{We recall that this means that there are no nontrivial (i.e. non-constant) meromorphic functions with poles only at the points of $\scr D$ and of degree not greater than the corresponding multiplicity of the point. Equivalently (by the Riemann--Roch theorem), there are no non-zero holomorphic differentials vanishing at the points of  $\scr D$ of the corresponding order.}
These data define uniquely a Cauchy kernel $\Cauchy(p,q)$ \cite{Fay, Zvero}. This is the unique function w.r.t. $q$ and meromorphic differential w.r.t. $p$ with the following divisor properties (the subscript refers to the variable for which the divisor properties are being assessed):
\bea
& (\Cauchy(p,q))_q\geq \infty - p-\scr D\nn \\
& (\Cauchy (p,q))_p\geq -\infty -q + \scr D.
\label{Cauchy}
\eea
and normalized by the requirement $\res{p=q}\Cauchy(p,q)=1 = -\res{p=\infty} \Cauchy(p,q)$.
\begin{example}
In genus $0$ by choosing $\infty$ as the point at infinity, the kernel takes the familiar form $\Cauchy(w,z) = \frac {\d w} {w-z}$. In this case $\scr D$ is the empty divisor.
In genus $1$, by representing the elliptic curve $\CC$ as the quotient $\C/ \Z + \tau \Z$, we can write it in terms of the Weierstra\ss\ $\zeta$ function; if $\scr D=(a)$ and we choose the point $\infty$ as the origin, for example,  
\be
\Cauchy(w,z) = \bigg(\zeta (w) - \zeta(w-z) - \zeta(a) + \zeta(a-z) \bigg)\d w
\ee
\end{example} 
One can write an expression  for $\Cauchy$ in terms of Theta functions; this can be found in more general setting in Section \ref{KPsect}. For hyperelliptic curves we give some really explicit expression in Section \ref{sectexample}.

Let $\gamma$ be a closed contour avoiding $\scr D, \infty$ and $\d\mu$ a smooth complex valued measure on it: with this we mean that in the neighbourhood of each point $p\in \gamma$, with $z$ a local coordinate in the neighbourhood, we can write $\d\mu(p) = f(z,\ov z) \d z$ where $f(z, \ov z):\gamma \to \C$ is a smooth function.

\begin{definition}
\label{DefWeyl}
The {\it Weyl}(Stiltjes) function of $\d\mu$ is the following differential on $\CC\setminus \gamma$:
\be
\label{defW}
W(p) = \int_{q\in \gamma} \Cauchy(p,q) \d \mu(q)
\ee
\end{definition}
We note that $(W)\geq \scr D-\infty$ on $\CC\setminus \gamma$ and that the residue at $\infty$ is simply the total mass of $\d\mu$ on $\gamma$.

We want to construct a Pad\'e--like approximation to  $W$ on $\CC$; in the standard setting $\CC= \P^1$ and $\scr D$ is empty and the Cauchy kernel is  $\Cauchy(z,w) = \frac {\d w}{w-z}$. Omitting the $\d w$, the Weyl function is really a function and not a differential: $W(w) = \int \frac {\d\mu(z)}{w-z}$. In this case the typical Pad\'e\ approximation problem is that of finding polynomials $P_n(x)$ of degree $\leq n$ and $Q_{n-1}$  of degree $\leq n-1$ such that 
\be
\label{pade0}
\frac {Q_{n-1}(z)}{P_n(z)} - W(z) =\mathcal O(z^{-(2n+1)}).
\ee
If we interpret the above equation as a statement about the vanishing at infinity of the meromorphic  differential $\frac {Q_{n-1}(z)}{P_n(z)}\d z$ we see that the order of vanishing is $2n-1$: the reader should not be confused here by the apparent discrepancy with the usual Pad\'e\ requirement that the order of vanishing is $2n+1$ because here we are considering the left side of \eqref{pade0} as a {\em differential} on $\P^1$ and $\d z$ has a double pole at infinity. 

We should then interpret the numerator as a meromorphic differential $Q_{n-1}\d z$ with a single pole at $\infty$ of order $n+1$.  

With this interpretation the Pad\'e\ problem can be similarly stated on $\CC$.
The main difference in the higher--genus case is that for a given measure $\d\mu$ there is a $g$--parametric family of Weyl functions parametrized by the choice of divisor $\scr D$. 

\begin{definition}[Pad\'e\ approximation]
\label{defPade}
Given $(\d\mu, \gamma,\scr D, \infty)$ as above, the $n$--th Pad\'e\ approximation is the datum of $P_n\in \scr L(\scr D+n\infty)$ and $\mathfrak Q_{n-1}\in \mathcal K((n+1)\infty)$ such that 
\be
\label{Pade}
\le(\frac {\mathfrak Q_{n-1}}{P_n} - W \ri)\geq 2\scr D+  (2n-1)\infty.
\ee
\end{definition}
We recall that the symbol $\scr L(\scr D+n\infty)$ denotes the vector space of meromorphic functions $f$ such that $(f)\geq - \scr D-n\infty$ and, similarly, the symbol  $\mathcal K((n+1)\infty)$ denotes the vector space of meromorphic differentials $\omega$ such that $(\omega)\geq  - (n+1)\infty$. Under our non-specialty assumption the Riemann--Roch theorem implies that generically the dimension of $\scr L(\scr D+n\infty)$ is $\dim\scr L(\scr D+n\infty)= n+1$. Similarly $\dim (\mathcal K((n+1)\infty)) = g+n$.

Let us draw some consequences from \eqref{Pade}; multiplying by $P_n$ the equation becomes 
\be
\le(\mathfrak Q_{n-1} - P_n W\ri) \geq \scr D+ (n-1)\infty.
\ee
Recalling the definition \ref{defW} of $W$ we can rewrite the above as follows:
\bea
\label{17}
\mathfrak Q_{n-1}(p) - \int_{q\in\gamma} (P_n(p)-P_n(q))\Cauchy(p,q) \d\mu(q) - \mathfrak R_n(p) = \mathcal O(-\scr D- (n-1)\infty)
\eea
where the remainder $\mathfrak R_n(p)$  is defined by 
\be
\label{defR}
\mathfrak R_n(p) = \oint_{q\in \gamma}\Cauchy(p,q) P_n(q) \d\mu(q).
\ee
%\blue
{The $\mathcal O$ notation above is used as follows: to say that $f=\mathcal O(\mathcal V)$ for a divisor $\mathcal V = \sum_{j} k_j p_j$, $p_j\in \CC$,  means that near each of the points $p_j$ the function (or differential) has a pole of order at most $k_j$ if $k_j>0$ and a zero of order at least $-k_j$ if $k_j<0$. We are following here the convention of algebraic geometry. Namely, in \eqref{17} the notation means that the differential vanishes at $\scr D$, and at infinity of order at least $n-1$. 
}
Note that the piecewise analytic differential  $\mathfrak R_n$ in \eqref{defR} has at most a simple pole at $\infty$  and vanishes at $\scr D$.

Since   we impose vanishing at $\infty$ to order $n-1$ on the right side of \eqref{17} we deduce that 
\be
\mathfrak Q_{n-1}(p) = \int_{q\in\gamma} (P_n(p)-P_n(q))\Cauchy(p,q) \d\mu(q).
\ee
which --we observe-- is a meromorphic differential on $\CC$ (there is no jump on $\gamma$)  with only one pole of order $n+1$ at $\infty$ ($n$ come from $P_n$ and $+1$ from the Cauchy kernel).
In principle one may want to add a differential $\omega_0$ to $\mathfrak Q_{n-1}$ to have a more general solution. However this $\omega_0$ should have at most one simple pole at $\infty$ (and hence no pole at all given that the sum of all residues must vanish) and moreover vanish at $\scr D$ since $\mathfrak R_n$ already vanishes there. By the property of non-special divisors recalled at the beginning of this section we conclude that $\omega_0\equiv 0$. 

Now, the Pad\'e\ approximation requires that the {\it remainder} term $\mathfrak R_n$ also  vanishes  at  $\infty$ of order $n-1$. 
 
Since it already vanishes at $\scr D$ by the definition of the Cauchy kernel, the extra requirements  give $n$ linear constraints on $P_n$ and hence generically we can expect  a unique solution. 
We investigate these conditions in the following sections.

\subsection{Pseudo moments and constructions of the Pad\'e\ approximants}

Let $z(p)$ be a local coordinate in the neighbourhood of $\infty$ such that $1/z(\infty)=0$ (i.e. mapping a punctured neighbourhood of $\infty$ to the outside of the unit disk). 

\bp
The following functions provide a  basis of $\scr L(\scr D+n\infty)$
\be
\label{zetaj}
\zeta_j(p) = -\res{q=\infty} z(q)^j \Cauchy(q,p), \qquad j=0,\dots, n
\ee
with the property  
\be
\zeta_j(p) = z(p)^{j}  + \mathcal O(z(p)^{-1}), \ \ \ \ p\to \infty.
\label{zetak}
\ee
\ep 
{\bf Proof.}
Given the divisor properties of $\Cauchy$ in \eqref{Cauchy}  it is evident that $\zeta_j$ has poles at $\scr D$ of the appropriate orders.
For the behaviour near $\infty$ we work in the local coordinate $z(p)$; let $z = z(p)$ and $w=z(	q)$. Then the residue  formula \eqref{zetaj} becomes
\be 
\oint_{|w|=R} w^j C(w,z)\d w
\ee
where orientation of the integration is counterclockwise and the Cauchy kernel can be written  $C(w,z) = \frac 1{w-z} + H(w,z)$ with $H(w,z)$ jointly analytic  in $z,w$ in the neighbourhood of $z=\infty=w$ and $H(w,z) = \mathcal O(1/z) \mathcal O(1/w^2)$. Then a simple application of Cauchy's residue theorem yields 
\be
\zeta_j = z^j +\mathcal O(1/z).
\ee
This immediately shows that $\zeta_j$ are linearly independent and span the required space of meromorphic functions. \QED

Note that $\zeta_0 \equiv 1$ since $\scr D $ is non-special.
Consider the coefficients $\mu_{n,j}$ defined by the following expansion:
\be
\label{exxp}
\oint_\gamma\Cauchy (p,q) \zeta_n(q)  \d \mu(q)  = \le(\sum_{j=0}^{\infty} \frac{\mu_{n,j}} {z(p)^j}  \ri)\frac{\d z(p)}{z(p)} .
\ee
We call them {\bf pseudo-moments} because in the case $\CC=\P^1$ they correspond to the usual moments of the measure. Note however that they do not form a Hankel matrix in general.
\bet
\label{propbilin}
{\bf [1]} The pseudo-moments $\mu_{j,k}$ in \eqref{exxp} are symmetric $\mu_{j,k}=\mu_{k,j}$ and can be written as 
\be
\label{hankel}
\mu_{j,k} = \oint_\gamma \zeta_j(p) \zeta_k(p) \d\mu(p)
\ee
{\bf[2]} More generally, for any two holomorphic  sections $\phi, \psi$ of $\scr L(\scr D)\bigg|_{\CC\setminus \{\infty\}}$ the following pairing 
\be
\label{bilmu1}
\big\langle \phi,\psi \big \rangle_\mu =- \res{q=\infty} \oint_{p\in\gamma}\phi(q)\Cauchy(q,p) \psi(p) \d \mu(p)
\ee
is symmetric and equals 
\be
\label{bilmu2}
\big\langle \phi,\psi \big \rangle_\mu =\oint_\gamma \psi(p)\phi(p)\d\mu(p).
\ee
\eet
\begin{remark}
In the genus zero case we have trivially $\zeta_j =z^j$ and the matrix of coefficients is a Hankel matrix. 
\end{remark}
\begin{remark}
In the second statement of Theorem \ref{propbilin} the wording simply means that $\phi,\psi$ are meromorphic functions on the {\em punctured} surface (i.e. at most with an isolated singularity at $\infty$) and such that their divisor of poles is bounded by $-\scr D$. We are mostly interested in the case when the singularity at $\infty$ is a pole of finite order, but the statement itself allows for functions with essential singularities.
\end{remark}
{\bf Proof.} {\bf [1]} 
Let 
\be
\phi_n(p):= \oint_\gamma \zeta_n(q) \Cauchy (p,q) \d \mu(q) .
\ee
This is a differential with a discontinuity across $\gamma$ and at most a simple pole at $p=\infty$. 
The coefficient $\mu_{n,j}$ can then be written as 
\be
\mu_{n,j} = \res{p=\infty} z(p)^j \phi_n(p) = \res{p=\infty} \zeta_j(p) \phi_n(p) 
\ee
where the second equality follows from the fact that $\zeta_j(p)- z(p)^j$ vanishes at $p=\infty$.
Rewriting this latter equality in terms of the definition of $\phi_n$ we have 
\be
\mu_{n,j} = -\res{p=\infty}\int_{q\in \gamma}  \zeta_j(p) \Cauchy(p,q)\zeta_n(q)\d\mu(q) =-\int_{q\in \gamma}  \res{p=\infty} \zeta_j(p) \Cauchy(p,q)\zeta_n(q)\d\mu(q) 
\label{munj}
\ee
where the last equality follows from Fubini's theorem because $\infty\not\in \gamma$.
Now we observe that the differential with respect to $p$ given by $\zeta_j(p)\Cauchy(p,q)$ has only poles at $p=\infty$ and $p=q$ (no poles at $p\in \scr D$ because of \eqref{Cauchy}) with opposite residues. Since $\res{p=q} \Cauchy(p,q)=1$ the Cauchy theorem shows that 
\be
-\res{p=\infty} \zeta_j(p) \Cauchy(p,q) = \zeta_j(q).
\label{ast}
\ee
Substituting \eqref{ast} into \eqref{munj} yields the proof.\\
{\bf [2]}  The equality of \eqref{bilmu1} and \eqref{bilmu2} is proved exactly as above and then the symmetry is evident in \eqref{bilmu2}.
\QED

The solution of the Pad\'e\ approximation problem \eqref{Pade} is then  predicated on the existence of $P_n\in \scr L(\scr D+n\infty)$ such that 
\be
\label{vanish}
\mathfrak R_n(p) = \int_\gamma \Cauchy(p,q)P_n(q)  \d \mu(q)  = \frac 1  {z(p)^{n+1}}\le(c + \mathcal O(z^{-1}) \ri) \d z(p).
\ee
If we write $P_n = \sum_{\ell=0}^n \pi_{n\ell} \zeta_\ell(p)$ the condition becomes that $\sum_{\ell=0}^n \pi_{n,\ell} \mu_{\ell,j}=0$ for $j=0,1,\dots, n-1$. In view of the Theorem \ref{propbilin} this can be written as 
\be
\le\langle P_n, \zeta_\ell\ri\rangle_\mu=0\ ,\ \ \ \ell=0,1,\dots, n-1.
\ee
which is the proxy of the usual orthogonality property for the ordinary Pad\'e\ approximants.

The study of the compatibility of the above system in the zero genus case is part of the theory of the Pad\'e\ table, \cite{Baker}, which is critically reliant upon the fact that the matrix of moments is a Hankel matrix. 
%
%We follow a different path by generalizing the characterization of orthogonal polynomials in terms of a Riemann--Hilbert problem \cite{FIK, FIK2}.

%
\subsection{Riemann--Hilbert problem}
\label{secRHP}
Like in the standard Fokas-Its-Kitaev \cite{FIK, FIK2} formulation of orthogonal polynomials, we can setup a Riemann--Hilbert problem on the Riemann surface $\CC$ which characterizes these Pad\'e\ denominators.

\begin{problem}
\label{RHP}
Let $Y_n$ be a $2\times 2$ matrix with functions in the first column and differentials in the second column, meromorphic in $\CC\setminus \gamma$ and admitting boundary values on $\gamma$ that satisfy the jump relation
\be
Y_n(p_+) = Y_n(p_-) \le[
\begin{array}{cc}
1 & 2i\pi\d \mu(p)\\
0 & 1
\end{array}
\ri],\qquad p\in \gamma.\label{jump}
\ee
In addition we require that  the matrix is such that it has poles at $\scr D$ in the first column and zeros in the second column, and also the following  growth condition at $\infty$:
\bea
Y_n(p) &= \le[
\begin{array}{cc}
\mathcal O(\scr D+n\infty) & \mathcal O(-\scr D- (n-1)\infty)\\
\mathcal O(\scr D+(n-1)\infty )&  \mathcal O(-\scr D- (n-2)\infty)
\end{array}
\ri].\label{growth}\\
Y_n(p) &= \le(\1 + \mathcal O(z(p)^{-1})\ri) \le[\begin{array}{cc}z^n(p) & 0\\
0 & \frac {\d z(p)}{z^n(p)}
\end{array}\ri]\ ,\qquad p\to \infty.
\label{normalization}
\eea
\end{problem}
%\sout{The $\mathcal O$ notation above is used as follows: to say that $f=\mathcal O(\mathcal V)$ for a divisor $\mathcal V = \sum_{j} k_j p_j$, $p_j\in \CC$,  means that the near each of the points $p_j$ the function (or differential) has a pole of order at most $k_j$ if $k_j>0$ and a zero of order at least $-k_j$ if $k_j<0$. We are following here the convention of algebraic geometry.}\red{[(Moved earlier)}

{
%\color{blue}
Exactly like in the genus zero case, the relevance of the RHP \ref{RHP} is that if a solution exists, then the $(1,1)$ entry provides the orthogonal section $P_n(p)$.  See  Theorem  \ref{thmDn} below.
}
\begin{example}[The case $n=0$.]
If $n=0$ we see that $(Y_0)_{11}$ must be the constant $1$ and $(Y_0)_{21}$ must vanish because it would be a meromorphic function with poles at $\scr D$ and a simple zero at $\infty$ (which is then identically zero thanks to the assumption of non-specialty of $\scr D$).
Then  the solution is given by 
\be
Y_0(p) = \le[
\begin{array}{cc}
1 & W(p)\\
0 & \res{q=\infty} \Cauchy (p,q) \d z(q)
\end{array}
\ri].
\ee
Note that the $(2,2)$ entry is the unique meromorphic differential with a single double pole at $\infty$ (normalized according to the choice of coordinate $z$) and zeros at $\scr D$.
\end{example}
\paragraph{Uniqueness of the solution: algebro-geometric approach.}
The determinant of $Y_n$ does not have a jump across $\gamma$ because the jump matrix in \eqref{jump} is of unit determinant. It is therefore a meromorphic differential: from the growth conditions \eqref{growth} it follows that it can only have a double pole at $\infty$:
\be
\Delta_n(p):= \det Y_{n}(p) \in \mathcal K(2\infty).
\ee
Since $\Delta_n(p)$ is a differential with a double pole, it must have $2g$ zeros (counting multiplicity). Therefore
the usual argument about the uniqueness of the solution to the problem \eqref{jump}, \eqref{growth} fails from the start because the matrix $Y_n^{-1}(p)$ has $2g$ poles. Indeed the usual reasoning would be to assume that $\wt Y_n$ is another solution to the same problem and then consider the ratio
\be
R_n(p):= \wt Y_n(p) Y_{n}^{-1}(p).
\ee
This matrix of functions does not have a jump across $\gamma$ and it is therefore a priori a matrix of meromorphic functions. If we could conclude immediately that they are --in fact-- holomorphic, the Liouville theorem would imply that they are constants  and $R_n$ is then the identity matrix because of the normalization condition \eqref{normalization}.

However, so far, we can only conclude that $R_n$ has poles at the $2g$ zeros of $\Delta_n$. We denote by $\scr T$ the divisor of zeros of $\Delta_n$ and call it the {\em Tyurin divisor}. 

Consider a row $\sigma(p)$ of $R_n(p)$; it is a meromorphic function such that $R_n(p) Y_n(p)$ is holomorphic at all points of $\scr T$; this allows us to interpret $\sigma(p)$ as a global holomorphic section of a vector bundle, $\scr E$ of rank $2$ and degree $2g$ described hereafter. 

For each $p_\alpha \in \scr T$ let $\DD_\alpha$ be a small disk covering the point $p_\alpha$ in such a way that these disks are pairwise disjoint; let $\DD_0$ be $\CC\setminus \scr T$. Then we define the vector bundle by the transition functions
\be
\sigma_\alpha(p) = \sigma_0(p) g_{\alpha 0} (p),\qquad g_{\alpha 0}(p):= Y_n(p)\bigg|_{\DD_{\alpha}}
\ee
Then we see that the row $\sigma$ of $R_n(p)$ is a holomorphic  section restricted to the trivializing set $\DD_0$  of the  above bundle. 

The Riemann--Roch theorem implies that generically such a bundle has only $2$ holomorphic sections; they are the sections such that their restriction to $\DD_0$ are the constant vectors ${\bf e}_1^t, {\bf e}_2^t$. 
This shows that generically the solution of the Riemann--Hilbert problem is unique.\\[3pt]

This reasoning is probably a bit mysterious for the reader accustomed to usual Pad\'e\ approximants: in the next section we clarify  the uniqueness in a completely elementary way which is much closer to usual methods of Pad\'e\ theory. This is accomplished in Theorem \ref{thmDn}.
\subsubsection{Genericity}

Define the determinant 
\be
\label{nothankel}
D_n:= \det \big[\mu_{a,b}\big]_{a,b=0}^{n-1} = \frac 1{n!} \int_{\gamma^n} \le(\det\big[\zeta_{a-1}(p_b)\big]_{a,b=1}^n\ri)^2 \prod_{j=1}^n\d\mu(p_j)
\ee
The second equality is an application of the Andr\'eief identity. 
We observe, and leave the verification to the reader, that a change of coordinate around $\infty$ from $z$ to $\wt z$ modifies these determinants only by a non-zero constant (the $n$--th power of the differential of the change of coordinate from  $1/z$ to $1/\wt z$ evaluated at $\infty$).

\begin{remark}
In the genus zero case the determinants \eqref{nothankel}  are Hankel determinants of the moments of the measure $\d \mu$.
\end{remark}

The following theorem is the higher genus counterpart of the characterization theorem for orthogonal polynomials in terms of a Riemann--Hilbert problem \cite{FIK}. Note, however, that there is a difference between the genus zero and higher genus cases: in genus zero the uniqueness and existence of the solution go hand-in-hand, namely if the solution exists, then it is unique. In higher genus the solution may exists but not unique, although generically it is unique. 

The next theorem shows that the  (existence+uniqueness) is  completely predicated upon the non-vanishing of a principal minor of the matrix of moments, much in the same way as in the genus zero case. However, it may happen that the determinant vanishes and yet we have a solution (not unique). This occurrence is precisely the non-vanishing of $h^1(\scr E)$ discussed above.

\bet
\label{thmDn}
If the determinant $D_n$  in \eqref{nothankel} does not vanish then the solution to the RHP \ref{RHP} exists and is unique. 
Viceversa if the solution exists and it is unique, then $D_n \neq 0$. 
Moreover the $(1,1)$ entry is the ``monic'' orthogonal section $P_n\in \scr L(n\infty + \scr D)$ of the form $P_n (p) = \zeta_n(p)+ \sum_{j=0}^{n-1} c_j \zeta_j(p)$. 
\eet
\noindent{\bf Proof.}
Suppose $D_n\neq 0$. 
Define, in a similar vein to the usual case of orthogonal polynomials, 
\be
P_n(p)  = \frac 1 {D_n} \det\le[
\begin{array}{cccccc}
\mu_{0,0} & \mu_{1,0} &\cdots & \mu_{n,0}\\
\mu_{0,1} & \mu_{1,1} &\cdots & \mu_{n,1}\\
\vdots &&&\vdots\\
\zeta_0(p) & \zeta_1(p) &\cdots & \zeta_n(p)
\end{array}
\ri].
\label{pn}
\ee
This is a section of $\scr L(\scr D+n\infty)$ of the form $P_n(p)= \zeta_n + \C\{\zeta_0,\dots, \zeta_{n-1}\}$ and hence  behaves as $z(p)^n$ as $p\to\infty$. Similarly we define 
\be
\wt P_{n-1}(p)  = \frac 1 {D_n} \det\le[
\begin{array}{cccccc}
\mu_{0,0} & \mu_{1,0} &\cdots & \mu_{n-1,0}\\
\mu_{0,1} & \mu_{1,1} &\cdots & \mu_{n-1,1}\\
\vdots &&&\vdots\\
\zeta_0(p) & \zeta_1(p) &\cdots & \zeta_{n-1}(p)
\end{array}
\ri] \in \scr L(\scr D+(n-1)\infty).
\ee
Finally we set 
\be
\mathfrak R_n(p):= \int_\gamma \Cauchy(p,q)P_n(q)\d\mu(q)\qquad
\wt {\mathfrak R}_{n-1}(p):= \int_\gamma \Cauchy(p,q)\wt P_{n-1}(q)\d\mu(q).
\ee
Consider then the matrix 
\be
Y_n(p):= \le[
\begin{array}{cc}
P_n (p)  & \mathfrak R_n(p)\\
\wt P_{n-1} (p)  & \wt{\mathfrak R}_{n-1}(p)
\end{array}
\ri].
\ee
A simple application of the Sokhostki-Plemelj formula shows that it satisfies \eqref{jump}. Near the divisor $\scr D$ it has the required growth in \eqref{growth} because of the properties \eqref{Cauchy} of the Cauchy kernel. It remains to verify the growth near $\infty$ and the normalization condition \eqref{normalization}.

The first column is clearly of the form $[z^n + \mathcal O(z^{n-1}), \mathcal O(z^{n-1})]^t$ and hence we need to focus only on the behaviour of the second column near $\infty$.

Consider the expansion of $\mathfrak R_n$ near $\infty$:
\be
\mathfrak R_n(p) =\le( \sum_{\ell=0}^\infty \frac{c_{\ell,n}}{z^\ell} \ri)\frac {\d z}{z}.
\ee
According to Theorem \ref{propbilin} we have 
\be
c_{\ell,n} =-\res{p=\infty} z(p)^\ell \mathfrak R_n(p) = -\res{p=\infty} \zeta_\ell(p) \mathfrak R_n(p) = \int_\gamma \zeta_\ell P_n\d\mu \mathop{=}^{\eqref{pn}} \frac 1 {D_n} \det\le[
\begin{array}{cccccc}
\mu_{0,0} & \mu_{1,0} &\cdots & \mu_{n,0}\\
\mu_{0,1} & \mu_{1,1} &\cdots & \mu_{n,1}\\
\vdots &&&\vdots\\
\mu_{0,\ell} & \mu_{1,\ell}&\cdots & \mu_{\ell,n}
\end{array}
\ri].
\ee
This expression clearly vanishes  for $\ell \leq n-1$ and hence indeed $\mathfrak R_n(p) = \mathcal O(z^{-n})  \frac{\d z}z$ near $\infty$.
In particular the leading coefficient of the expansion is 
\be
\label{leadcof}
\mathfrak R_n(p) =  \frac {D_{n+1}}{D_n} \frac 1{z^{n+1}}\le(1 + \mathcal O(z^{-1}) \ri) \d z. 
\ee
The same computation for $\wt {\mathfrak R_n}$  gives that 
\be
\wt {\mathfrak R_n}(p) = \frac 1 {z^n} (1 + \mathcal O(z^{-1}))\d z
\ee
which satisfies the growth condition \eqref{growth} and the normalization \eqref{normalization} as well. \\[10pt]
Having shown the existence, we now need to address the uniqueness of the solution. 
Let $\wt Y$ (we omit the subscript $_n$ for brevity) be a solution of RHP \ref{RHP}:
the jump condition \eqref{jump} implies that the first column of the solution must be made of sections of $\scr L(\scr D+n\infty)$ and $\scr L(\scr D+(n-1)\infty)$. The same jump condition implies that the second column is obtained from the first by the integral against the Cauchy kernel: this is so because the divisor $\scr D$ is non-special and there is no nontrivial holomorphic differential that vanishes at $\scr D$. 

Next, the order of vanishing at $\infty$ of $\wt Y_{12}$ must be $n-1$ (i.e. it must be of the form $\mathcal O(\frac {1}{z^{n+1}})\d z$); the same computation used above implies then that 
\be
\wt Y_{11} (p) = P_n(p) + \sum_{\ell=0}^{n-1} \alpha_\ell \zeta_\ell(p)% = P_n(p) +S_{n-1}(p)
\ee
for $P_n$ as in \eqref{pn} and some coefficients $\alpha_\ell$. These coefficients must satisfy the linear system 
\be
\le[
\begin{array}{ccc}	
\mu_{00} & \dots& \mu_{0,n-1}\\
\vdots&&\\
\mu_{n-1,0} & \dots  & \mu_{n-1,n-1}
\end{array}
\ri] \vec \alpha =\vec 0 \ ,
\label{136}
\ee
which has only the trivial solution because of the assumption $D_n\neq 0$. 
Next, the component $\wt Y_{21}\in \scr L(\scr D+(n-1)\infty)$ is subject to similar constraints: writing it as a linear combination $\sum_{\ell=0}^{n-1} \beta_\ell \zeta_\ell$ we see that the asymptotic constraint that $\int_\gamma \Cauchy(p,q) \wt Y_{21}(q)\d\mu(q) = \frac 1 {z^n} (1+\mathcal O(z^{-1})\d z$ translates in the linear system:
\be
\le[
\begin{array}{ccc}	
\mu_{00} & \dots& \mu_{0,n-1}\\
\vdots&&\\
\mu_{n-1,0} & \dots  & \mu_{n-1,n-1}
\end{array}
\ri] \vec \beta =\le[
\begin{array}{c}
0\\
\vdots\\ 0\\1
\end{array}
\ri],
\ee
which, again, has a unique solution thanks to the assumption $D_n\neq 0$. \\[10pt]
We now show the converse statement. 
{
%\color{blue}
Suppose a solution $Y$ exists and is unique. Denote by ${\bf A}, {\bf B}$ the two columns of $Y$.
The jump condition \eqref{jump} together with the growth condition \eqref{growth} at the divisor $\scr D$ implies that
\begin{enumerate}
\item[(i)] ${\bf A}$  is meromorphic on $\mathcal C$ and the first entry, $A_1(p)$, is a section in $\scr L(\scr D +n\infty)$;
\item[(ii)] ${\bf B}(p) = \int_\gamma \Cauchy(p,q) {\bf A}(q) \d \mu(q)$. 
\end{enumerate}
From \eqref{normalization} it follows that the entry $A_1(q)$ has a pole of order exactly $n$ at $\infty$ and hence it can be written as a linear combination 
\be
A_1(p) = \sum_{\ell=0}^{n} c_\ell \zeta_\ell(p),\ \ \ c_n=1.
\ee
The vanishing of order $n-1$ at $\infty$ of the first entry $B_1(p) = \int_\gamma\Cauchy(p,q) A_1(q) \d\mu(q)$ is equivalent to the condition  that the coefficients $c_0,\dots, c_{n}$ satisfy the system 
\be
\le[
\begin{array}{ccc}	
\mu_{00} & \dots& \mu_{0,n}\\
\vdots&&\\
\mu_{n,0} & \dots  & \mu_{n,n}
\end{array}
\ri] \vec c =\le[\begin{array}{c}
{\ds 0\atop
{\vdots\atop \ds 0}}
\\
\star
\end{array}\ri] \ , \label{csys}
\ee
 where $\star$ denotes a possibly nonzero coefficient. 

So far we have used only the existence of the solution; now we use the uniqueness assumption to show that $D_n\neq 0$.   Suppose, by contrapositive, that $D_n=0$ and let $(\alpha_0,\dots, \alpha_{n-1})$ be a nontrivial solution of \eqref{136}. Then the vector $\vec {\wt c}= (c_0+\alpha_0, \dots, c_{n-1}+\alpha_{n-1}, 1)$ is another solution of the same equation \eqref{csys}, which then violates the uniqueness.} \QED

The Theorem \ref{propbilin} allows us to interpret the vanishing condition of $\mathfrak R_n$ \eqref{leadcof} precisely as an ``orthogonality'' 
\be
 \int_{\gamma} P_n(p)\zeta_j(p)\d\mu(p) = \delta_{jn} \frac {D_{n+1}}{D_n}\ ,\ \ \forall j\leq n.
\ee
This latter equation, in turn implies 
\be
 \int_{\gamma} P_n(p)P_j(p)\d\mu(p) = \delta_{jn} \frac {D_{n+1}}{D_n}.
\ee
If all the sequence of determinants $\{D_n\}_{n\geq 0}$ does not vanish, the above condition is then the usual (non-hermitean) orthogonality.

\paragraph{Existence without uniqueness.}
Suppose that $D_n=0$; the solution to the RHP \ref{RHP}  may still exist. For this to happen  we must  find a linear combination $P_n(p) = \zeta_n(p)- \sum_{\ell=0}^{n-1} \alpha_\ell \zeta_\ell(p)$ which is orthogonal to $\zeta_j, \ j=0,\dots, n-1$. 
Moreover there must be also a $\wt P_n\in \scr L(\scr D+(n-1)\infty)$ such that its Cauchy transform is appropriately normalized.
This may happen if both the following systems are simultaneously compatible:
\be
\le[
\begin{array}{ccc}	
\mu_{00} & \dots& \mu_{0,n-1}\\
\vdots&&\\
\mu_{n-1,0} & \dots  & \mu_{n-1,n-1}
\end{array}
\ri] \vec \alpha =\le[
\begin{array}{c}
\mu_{0,n}\\
\vdots\\
\mu_{n-1,n}
\end{array}
\ri],\qquad
\le[
\begin{array}{ccc}	
\mu_{00} & \dots& \mu_{0,n-1}\\
\vdots&&\\
\mu_{n-1,0} & \dots  & \mu_{n-1,n-1}
\end{array}
\ri] \vec \beta =\le[
\begin{array}{c}
0\\
\vdots\\
1
\end{array}
\ri],
\label{245}
\ee
where $\wt P_n(p) = \sum_{\ell=0}^{n-1} \beta_\ell \zeta_\ell(p)$. 
The expression 
\be
Q_{n-1}(p):= \det\le[
\begin{array}{cccccc}
\mu_{0,0} & \mu_{1,0} &\cdots & \mu_{n,0}\\
\mu_{0,1} & \mu_{1,1} &\cdots & \mu_{n,1}\\
\vdots &&&\vdots\\
\zeta_0(p) & \zeta_1(p) &\cdots & \zeta_n(p)
\end{array}
\ri] 
\ee
belongs to $\scr L(\scr D+(n-1)\infty)$ (the coefficient in front of $\zeta_n$ vanishes in the Laplace expansion) and has also the property that its Cauchy transform vanishes at $\infty$ like $\d z/z^{n+1}$ since it is orthogonal to all $\zeta_0, \dots, \zeta_{n-1}$. Thus the row--vector 
\be
\sigma_n(p):= \le[Q_{n-1}(p), \int_\gamma \Cauchy(p,q)Q_{n-1}(q)\d\mu(q)\ri]
\ee 
is a row--vector solution that can be added to either rows of $Y_n$ and the uniqueness of the solution is lost. \\[1pt]

{
%\color{blue}
 Note  that in genus zero  $\mu_{a,b} = f_{a+b}$ is a Hankel matrix and the vanishing of $D_n$ makes the two systems \eqref{245} incompatible. A simple  way to convince ourselves of this fact is to count the number of constraints versus the number of equations. 
Indeed, if $D_n=0$ (viewed as a polynomial relation in the coefficients), the compatibility of the two systems imposes additional $2n$ equations for the $2n$ indeterminates $f_0,\dots, f_{2n-1}$; these additional equations are given by the vanishing of all the determinants obtained by replacing the $n$ columns of the Hankel matrix by either of the two vectors on the right sides of \eqref{245}. Thus in the end one has $2n+1$ polynomial equations in $2n$ variables. This argument, while simple and  perhaps convincing, is not entirely satisfactory because the $2n+1$ polynomial relations should be proved algebraically independent.

A complete but indirect  proof is obtained by invoking the fact that, in genus zero, the existence of the solution to the Riemann-Hilbert problem implies automatically its uniqueness. Indeed the system \eqref{245} provides the solution by setting $Y_{11}(z) = z^n + \sum_{j=0}^{n-1} \alpha_j z^j,\ \ Y_{21}(z) = \sum_{j=0}^{n-1} \beta_j z^j$ and defining the second column as $Y_{\bullet 2}(z) = \int_\gamma \frac {Y_{\bullet 1}(w)\d \mu(w)\d w}{z-w}$. If we assume by contrapositive that the two systems \eqref{245} are compatible under the assumption $D_n=0$, we obtain a contradiction with the uniqueness of the solution of the Riemann--Hilbert problem since neither $\vec \alpha$ nor $\vec \beta$ are uniquely defined. 
}

\paragraph{The case $\infty\in \gamma$ or $\scr D\cap \gamma \neq \emptyset$.}
We have assumed, for simplicity, that $\infty$ does not belong to the support $\gamma$ of the measure $\d\mu$. 
We can lift this restriction easily without modifying any of the substance. In this case we must assume that 
the functions $z^\ell$ are locally integrable at $\infty$ with respect to the measure $\d\mu$, for all $\ell\in \mathbb N$. 
Some modification in the statements about the growth  then needs to be made but it is of the same nature as the case of ordinary orthogonal polynomials. Similarly, if a  point $p_0$  of the divisor $\scr D$ (of multiplicity $k_0$) belongs to $\gamma$ we need to  assume  that the function $1/\kappa^{2k_0}$ (with $\kappa$ a local coordinate at $p_0$) is locally integrable in the measure $\d\mu$ at $p_0$. Some technical considerations will have to be modified accordingly but the essential picture remains the same.

\paragraph{Heine formula.}
In the genus zero case the orthogonal polynomials can be expressed in terms of a multiple integral that goes under the name of Heine formula \cite{SzegoBook}. The following simple proposition expresses the orthogonal sections in a similar fashion. 
\bp[Heine formula]
\label{propheine}
The following section $\Psi_n(p)\in \scr L_n:= \scr L (\scr D+n\infty)$ is orthogonal to $ \scr L_{n-1}$:
\be
\Psi_n(p) := \int_{\gamma^n} \det\Big[\zeta_{a-1}(p_b)\Big]_{a,b=1}^{n+1}\det\Big[\zeta_{a-1}(p_b)\Big]_{a,b=1}^{n} \prod_{j=1}^n \d\mu(p_j), \qquad p_{n+1}= p.
\ee
If $D_n\neq 0$ in \eqref{nothankel} then the ``monic''  orthogonal section $P_n = \zeta_n + \dots \in \scr L_n \setminus\scr L_{n-1}$ defined in \eqref{pn}  is given in terms of $\Psi_n$  by $P_n(p) = \frac 1{n! D_n} \Psi_n(p)$.
\ep 
{\bf Proof.}
Let $j\leq n-1$ and consider $\int_\gamma \Psi_n(p) \zeta_j(p) \d \mu(p)$.
Using the Laplace expansion we have 
\be
\Psi_n(p) = \sum_{\ell=0}^{n} (-1)^{n-\ell} \zeta_\ell(p) \int_{\gamma^n} \det\Big[\zeta_{a-1}(p_b)\Big]_{b\in [1..n]\atop
a\in [1..n+1]\setminus\{\ell\} }\det\Big[\zeta_{a-1}(p_b)\Big]_{a,b=1}^{n} \prod_{j=1}^n \d\mu(p_j)
\ee
Using the Andreief identity we obtain
\be
\Psi_n(p) = n!\sum_{\ell=0}^{n} (-1)^{n-\ell} \zeta_\ell(p)
\det \Bigg[\mu_{j,k}\Bigg]_{j\in [1..n]\setminus\{\ell\}\atop k\in[1..n]}.
\ee
The latter expression is the Laplace expansion of the determinant 
\be
n!\det \le[\begin{array}{cccc}
\mu_{00}& \cdots & \mu_{0n}\\
\vdots & & \vdots\\
\mu_{n-1,0} &\cdots & \mu_{n-1,n}\\
\zeta_0(p) & \cdots & \zeta_n(p)
\end{array}\ri].
\ee
At this point it is clear that the expression is orthogonal to $\zeta_j, \ j=0\dots, n-1$, spanning $\scr L(\scr D+(n-1)\infty)$.
If $D_n\neq 0$ then $\Psi_n$ has actually a pole of order $n$ and can be ``normalized'' to be monic.
\QED

\subsection{Example: the hyperelliptic case}
\label{sectexample}
Let $\mathcal C$ be a hyperelliptic curve  of the form 
\be
y^2 = \prod_{j=1}^{2g+2}(z-t_j)
\label{hyper}
\ee
where the numbers $t_j$ are pairwise distinct. This is a Riemann surface of genus $g$ (compactified by adding two points above $z=\infty$). The reader may visualize it as a two--sheeted cover of the $z$--plane, branched at the points $t_j$'s. A simple way of doing so is to glue two copies of the $z$--plane dissected along pairwise disjoint segments joining the branchpoints in pairs (for example $[t_1,t_2]$, $[t_3,t_4]$ etc.)

A point $p\in \mathcal C$ is a pair of values $p=(z,y)$ satisfying the equation \eqref{hyper}. It is well known \cite{Farkas} that a degree $g$ non-special divisor is any divisor $\scr D = \sum_{\ell=1}^g p_\ell$ (points may be repeated) as long as $p_\ell = (z_\ell, y_\ell)$ are such that $y_\ell \neq -y_k, \ \ \ell\neq k$. Note that the points $p_\ell$ may be equal to one of the branch--points $t_\ell$'s but then it must be of multiplicity  one. For added simplicity in this example we assume that $z_j\neq z_k,\ j\neq k$.

We choose $\infty$ to be the point $z=\infty$ and on the sheet where $y(z)/z^{g+1}\to 1$: we denote this point as $\infty_+$, whereas $\infty_-$ is the point $z=\infty$ where $y(z)/z^{g+1}\to -1$.  

A simple exercise shows that the Cauchy kernel subordinated to the choice of divisor $\scr D$ and with pole at $\infty_+$ is given by  (here $p=(w,y(w))$ and $q=(z,y(z))$)
\be
\Cauchy(p,q) = 
\Bigg(\frac{ y(z)+y(w) }{w-z}+ \overbrace{\prod_{k=1}^g (w-z_k)}^{=:L(w)} +\sum_{\ell=1}^g L_\ell(w)\frac {y(z) + y_\ell }{z-z_\ell} \Bigg)  \frac {\d w}{2y(w)}
\ee
where $L_\ell(w)$ are the elementary Lagrange interpolation polynomials:
\be
L_\ell(w) = \prod_{k\neq \ell} \frac {w-z_k}{z_{\ell}-z_k}
\ee
To verify that this is the correct Cauchy kernel, one has to verify the divisor properties \eqref{Cauchy}: the least obvious might be the vanishing, as a function of $z$, when $q=(z,y(z))$ tends to $\infty_+$. 

This can be seen as follows: the part that does not obviously vanish comes from the term
\be
y(z) \le(\frac 1{w-z} + \sum_{\ell=1}^g\frac{ L_\ell(w) }{z-z_\ell}\ri) + L(w).
\label{LL}
\ee
Expanding the bracket in \eqref{LL}  in  geometric series w.r.t $z$   we have 
\be
 \frac 1 z \sum_{k=0}^\infty \frac 1 {z^k} \le(
-w^k + \sum_{\ell=1}^g L_\ell(w) z_\ell^k 
\ri) =- \frac{\prod_{\ell=1}^g(w-z_\ell)}{z^{g+1}}\le (1 + \mathcal O(z^{-1})\ri).
\label{254}
\ee
The last equality is due to the fact that, for $k\leq g-1$ the polynomial of $w$ in the bracket has degree $\leq g-1$ and vanishes at the $g$ points $z_1,\dots, z_g$; for $k=g$ it is a polynomial of degree $g$ with leading coefficient $-1$ and vanishing at all the $g$ points, so that necessarily equals to $-L(w)= -\prod_{\ell=1}^g(w-z_\ell)$. Multiplying \eqref{254} by $y(z)$ we see that the expression \eqref{LL} vanishes at $\infty_+$.

A  basis of functions $\zeta_j$ such that $\scr L(\scr D+n\infty) = {\rm Span}\{\zeta_0,\dots, \zeta_n\}$ can be taken to be
\be 
\zeta_0 = 1;\qquad
\zeta_j = \frac 1 2\le[\frac { z^{j-1} y(z)}{\prod_{\ell =1}^g (z-z_\ell)}+P_j(z) + \sum_{\ell=1}^g \frac { z_\ell^{j-1} y_\ell}{ (z-z_\ell) \prod_{k\neq \ell}(z_k-z_\ell)}\ri]
\ \ j\geq  1 
\ee
where $P_j(z)$ is the polynomial given by 
\be
P_j(z) =  \le(\frac { z^{j-1} y(z)}{\prod_{j=1}^g (z-z_j)}\ri)_+
\ee
with the subscript indicating the polynomial part and the determination of $y(z)$ being the one such that $y(z)/z^{g+1}\to 1$.
\subsubsection{Curves with antiholomorphic involution}
\label{antihol}
These are curves with an antiholomorphic diffeomorphism $\nu:\CC\to \CC$. Without entering into details, in the case of hyperelliptic curves above, these are curves such that the set of  branch-points  $\{t_1,\dots, t_{2g+2}\}$ is invariant under complex conjugation, and in this case the map $\nu$ is the map $\nu(z,y) = (\ov z, \ov y)$ (or $\nu(z,y)= (\ov z,-\ov y)$). In general, for a plane algebraic curve defined as the polynomial equation $E(z,y)=0$ this means that all the coefficients of $E$ are real.

If we choose also the divisor $\scr D$ invariant under $\nu$ and $\infty$ as a fixed point (in our case both points at $z=\infty$ are fixed by the map $\nu$), then the Cauchy kernel is also a real--valued kernel in the sense that $\Cauchy(\nu(p),\nu(q)) = \ov{\Cauchy(p,q)}$. The basis of sections $\zeta_j$ of $\scr L$ is then  real--valued as well so that $\zeta_j(\nu(p))=\ov{\zeta_j(p)}$.

We can then choose  $\gamma$ to be a closed contour fixed by $\nu$, $\nu(p)=p\, \ \ \forall p\in \gamma$  and $\d\mu$ a positive real--valued measure on $\gamma$. In this case the determinants $D_n$ \eqref{nothankel} will be strictly positive and hence the Theorem \ref{thmDn} shows that the solution of the RHP \ref{RHP} exists and is unique for all $n\in \mathbb N$. Therefore we have an infinite basis of orthogonal functions exactly as in the usual case of orthogonal polynomials for an $L^2(\R, \d\mu)$.
\paragraph{Genus $1$.}
An example where we can write in great details the objects described above  is the case of an  elliptic curve $E_\tau$  realized as quotient of the plane $\C$ by the lattice $2\omega_1\Z+ 2\omega_2 \Z$, with $\tau= \omega_2/\omega_1 \in i\R_+$. Without loss of generality, we can choose $\omega_1\in \R$. 
In Weierstra\ss\ form the elliptic curve is 
\be
Y^2 = 4X^3 - g_2 X-g_3 = 4(X-e_1)(X-e_2) (X-e_3)
\label{ellcurve}
\ee
 with $e_1+e_2+e_3=0$ and $e_1<e_2<e_3$ (all real) or $e_1=\ov e_2$, $e_3\in \R$. 
 For definiteness we consider the case where $e_1,e_2,e_3\in \R$: then $\omega_1 = \int_{e_1}^{e_2} \frac {\d X}{Y}$ (with $Y= \sqrt{4X^3 - g_2 X-g_3}$ chosen so that it is positive in $[e_1,e_2]$) and  the Weierstra\ss\  functions provide the uniformization of \eqref{ellcurve}. Setting 
\be
\zeta(z) =  \frac 1 {z} + \sum_{\ell,k\in \Z\atop (\ell,k)\not=(0,0)} \le(\frac 1 {(z+2\omega_1\ell + 2\omega_2 k)} - \frac 1{(2\omega_1\ell +2\omega_2 k)} - \frac z{(2\omega_1\ell +2\omega_2 k)^2} \ri),
\ee
then the Weierstra\ss\ $\wp$ function is $\wp=-\zeta'$:
\be
\wp(z) = \frac 1 {z^2} + \sum_{\ell,k\in \Z\atop (\ell,k)\not=(0,0)} \le(\frac 1 {(z+2\omega_1\ell + 2\omega_2 k)^2} - \frac 1{(2\omega_1\ell +2\omega_2 k)^2} \ri).
\ee
The classical result of uniformization is then obtained by setting $X = \wp$ and $Y = \wp'$. 

The resulting elliptic curve $E_\tau$ admits the obvious antiholomorphic involution $z\to \frac {\omega_1}{\overline {\omega_1}} \ov z= \ov z$.  We choose $\infty =\{0 \}$ and $\scr D=\{ a\}$, with $a\in (0,2\omega_1)$. 
A basis of sections of $\scr L(\scr D+n\infty)$ is provided in terms of the Weierstra\ss\ $\zeta$ and $\wp$ functions
\be
\scr L(\scr D+n\infty) = \C\le\{1,\zeta\le(z \ri)- \zeta\le(z-a\ri) - \zeta \le(a\ri) , \wp(z), \wp'(z), \dots, \wp^{(n-2)}(z)\ri \}.
\ee
Note that all these functions are real--analytic: $\ov{f(z)} = f(\ov z)$.
The Cauchy kernel is given by 
\be
\Cauchy(z,w) = \le(\zeta(z-w) + \zeta\le(w-a\ri) - \zeta(z) + \zeta(a) \ri)\d z.
\ee

As for contour of integration $\gamma$ we choose the $a$--cycle, which is represented as either $[\omega_1, \omega_1+2\omega_2]$ in the $z$--plane  or the segment $[e_2,e_3]$ in the $X$--plane, on both sheets of the curve. Note that $\ov \gamma = \gamma \mod \Z + \tau \Z $ (pointwise)

%\subsubsection{The case of $\scr D = \infty$}
%In this case the basis of sections is simpler and we can write even more explicit formul\ae.
%Indeed we can choose  
%\be
%\zeta_0(z) \equiv 1,\   \zeta_{k} = \wp^{(k-1)}\le(z-\frac{\tau}2 \ri), \ \ k\geq 1
%\ee
%so that $(\zeta_k)\geq -(k+1)\infty$. 

The simplest case of Weyl differential is for the flat measure $\d\mu(x) = \d x$ on $[\omega_2,\omega_2 + 2\omega_1]$, thought of as the $a$--cycle on the elliptic curve $E_\tau$. 
The Cauchy kernel is given by 
The Weyl differential $W(p)$ is then 
\be
W(z) = \int_{\omega_2}^{\omega_2+2\omega_1} \Cauchy(z,w)\d w
= \big(2\omega_1\zeta(z-a) + 2\omega_1 \zeta(a) - 2\eta_1 z \big)\d z 
\ee
where $\eta_j = \zeta(\omega_j)$ are Weierstra\ss\ eta functions (not to be confused with Dedekind's $\eta$ function) and satisfying the  Lagrange identity
\be
\label{Lagrange}
\eta_1  \omega_2  - \eta_2\omega_1 = \frac {\pi i}2.
\ee
The identity \eqref{Lagrange} implies, as it should be, that $W(z)$ has a jump--discontinuity along the segment $(0,2\omega_1)$ and its translates thanks to the quasi--periodicity of the $\zeta$ function 
\be
\zeta(z + 2\omega_j) = 2\eta_j.
\ee
Namely: $W(z +2\omega_2) - W(z) = 2i\pi \d z$ (as it should be from the definition).

The matrix of moments is almost a Hankel matrix because 
\be
\mu_{2+j,2+k} = \int_{\omega_2}^{\omega_2+2\omega_1} \wp^{(j)}(z) \wp^{(k)}(z) \d z  = (-1)^k\int_{\omega_2}^{\omega_2 + 2\omega_1} \wp^{(j+k)}(z) \wp(z) \d z.
\ee
In particular the integral is zero if $j,k$ have different parity. 
This latter integral in principle can be computed in closed form; indeed since the integrand is an elliptic function with only one pole, it can be written as a linear combination of $\wp^{(\ell)}$, $\ell=0,\dots, j+k$. Only the coefficient of $\wp$ in this latter expansion survives because all other terms integrate to zero. 
Using the well known formula for the expansion of $\wp$
\bea
\wp(z) &= \frac 1 {z^2} + \sum_{\ell =2}^\infty c_\ell z^{2\ell-2}, \ \ \ \cr
&c_2= \frac {g_2}{20},\ c_3 = \frac {g_3}{28}, \ \ c_\ell = \frac 3{(2\ell+1)(\ell-3)}\sum_{m=2}^{\ell-2} c_m c_{\ell-m},
\eea
 one finds easily 
\be
\mu_{2+j,2+k}= 2c_{(j+k)/2}\frac{(j+k+2)!}{j+k+1}\eta_1,
\ee
when $j,k$ have the same parity (and zero otherwise).
This is not of much use at any rate because there are no closed formulas for $\mu_{1,j}$. 
We note only that the first two rows and columns of the moment matrix do not satisfy the Hankel property. For example
\bea
\mu_{0,2+k} = \int_{\omega_2}^{\omega_2+ 2\omega_1} \wp^{(k)}(z)\d z =  (\zeta(\omega_2)-\zeta(\omega_2+2\omega_1))\delta_{k0} =-  2\eta_1\delta_{k0}.
\eea

\begin{figure}
\includegraphics[width=0.24\textwidth]{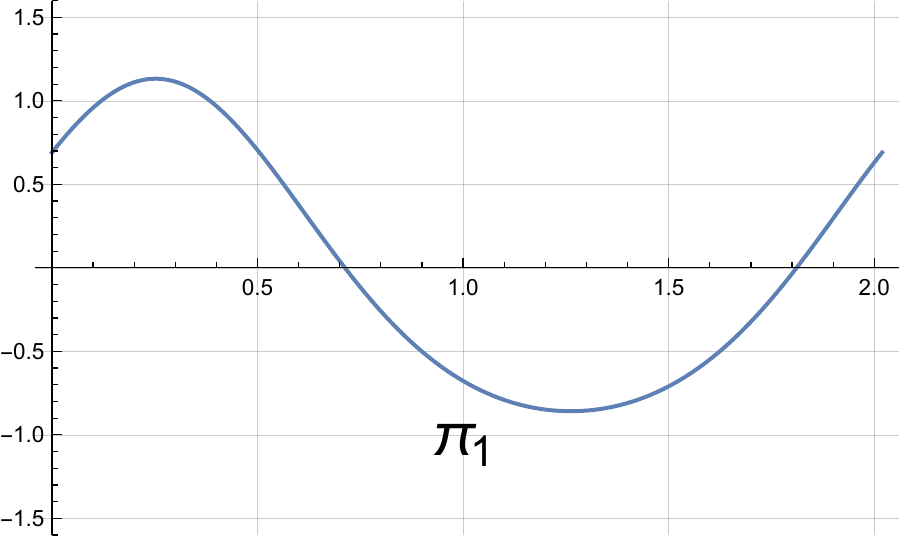}
\includegraphics[width=0.24\textwidth]{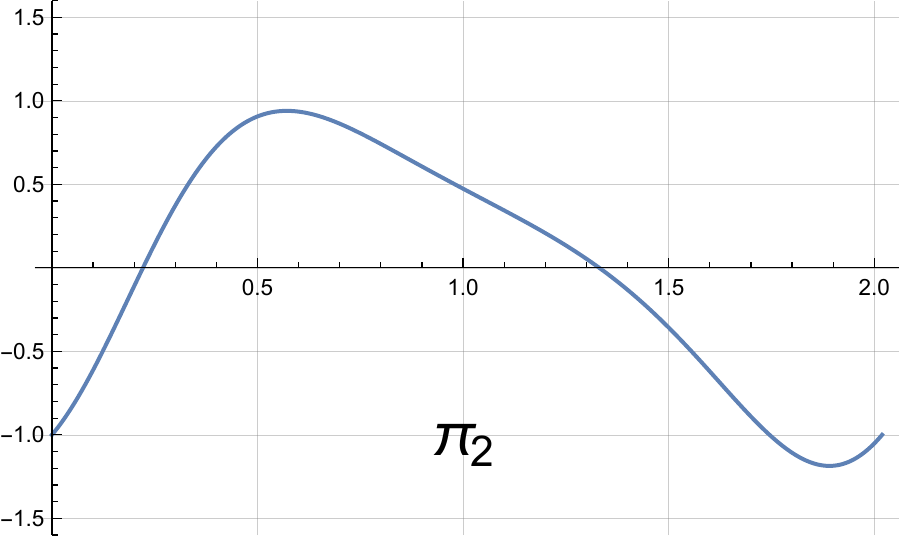}
\includegraphics[width=0.24\textwidth]{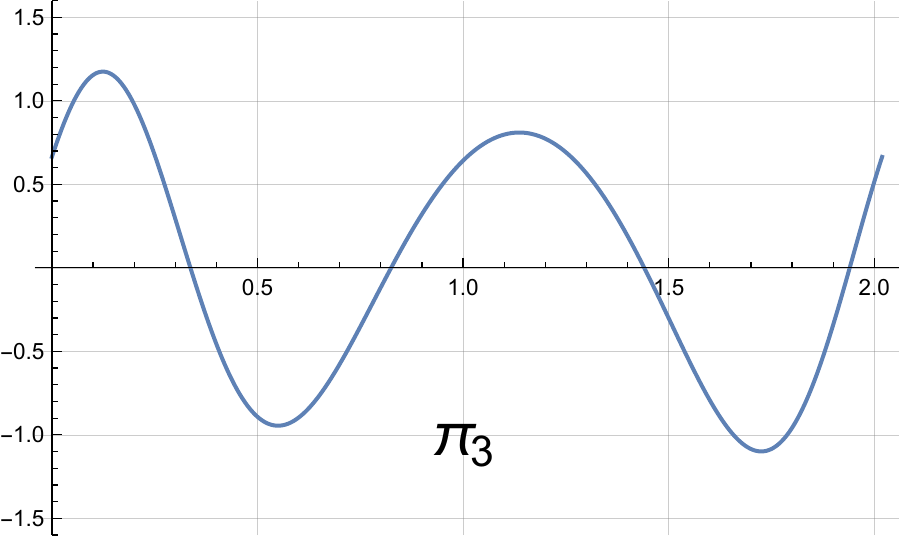}
\includegraphics[width=0.24\textwidth]{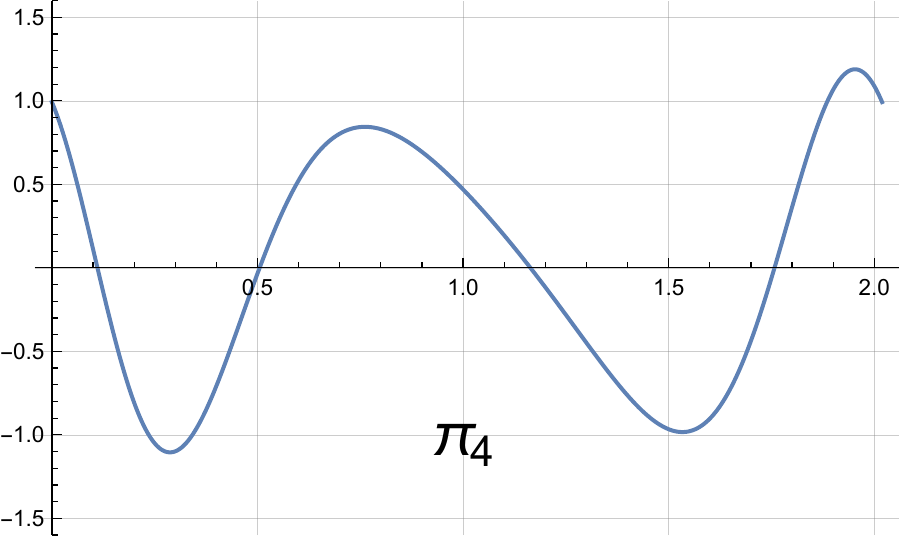}\\
\includegraphics[width=0.24\textwidth]{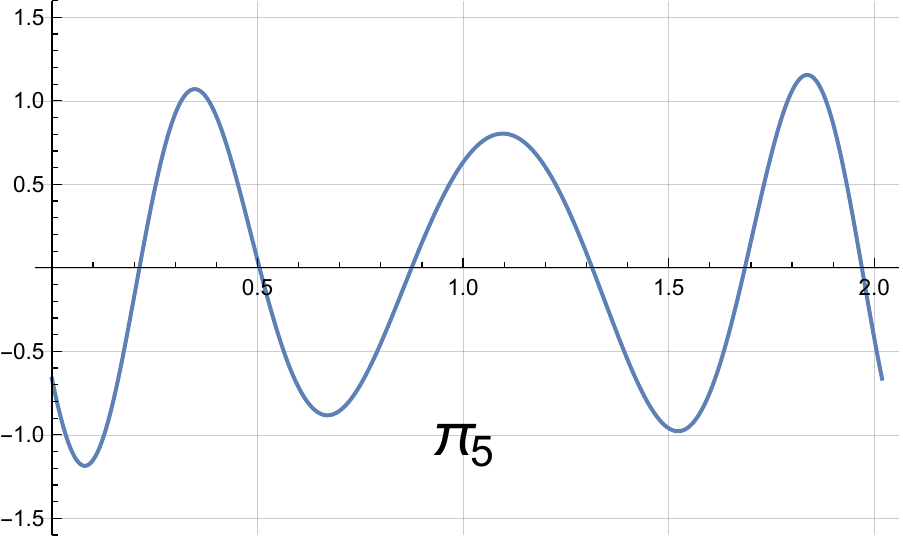}
\includegraphics[width=0.24\textwidth]{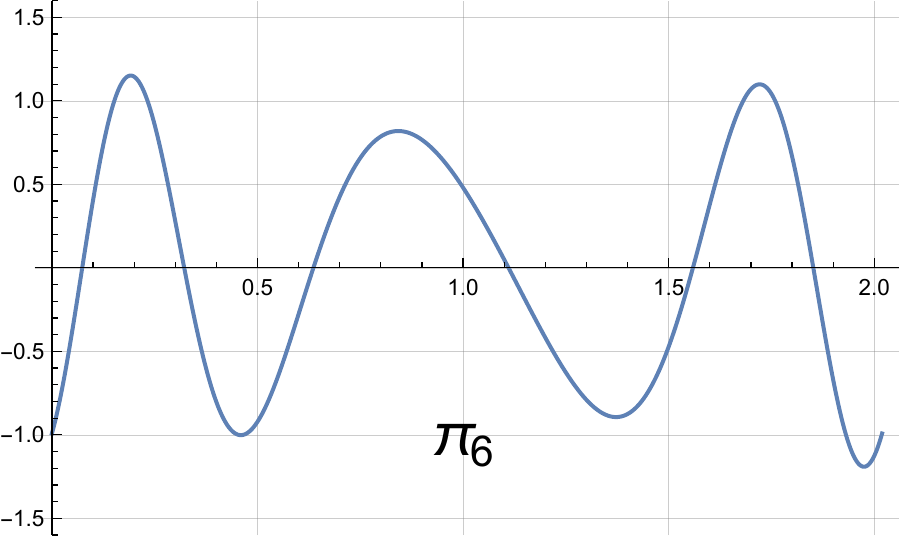}
\includegraphics[width=0.24\textwidth]{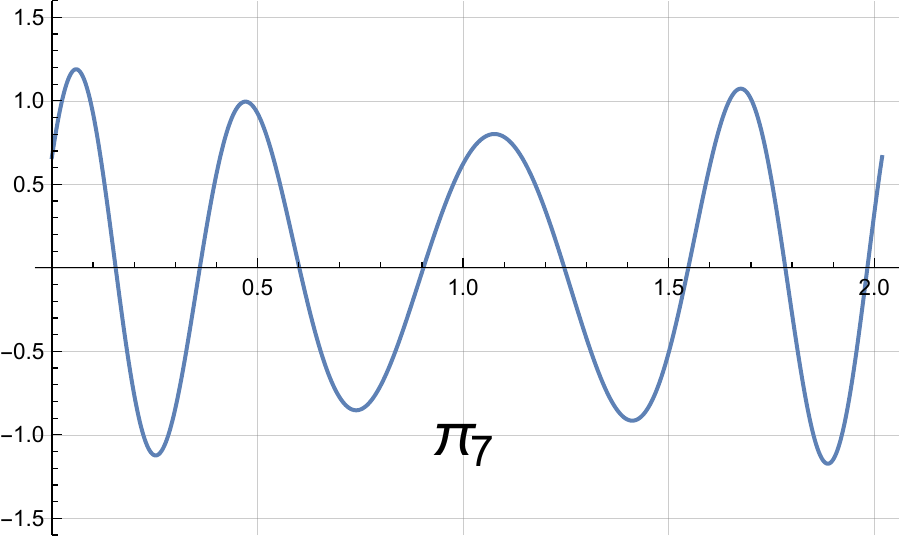}
\includegraphics[width=0.24\textwidth]{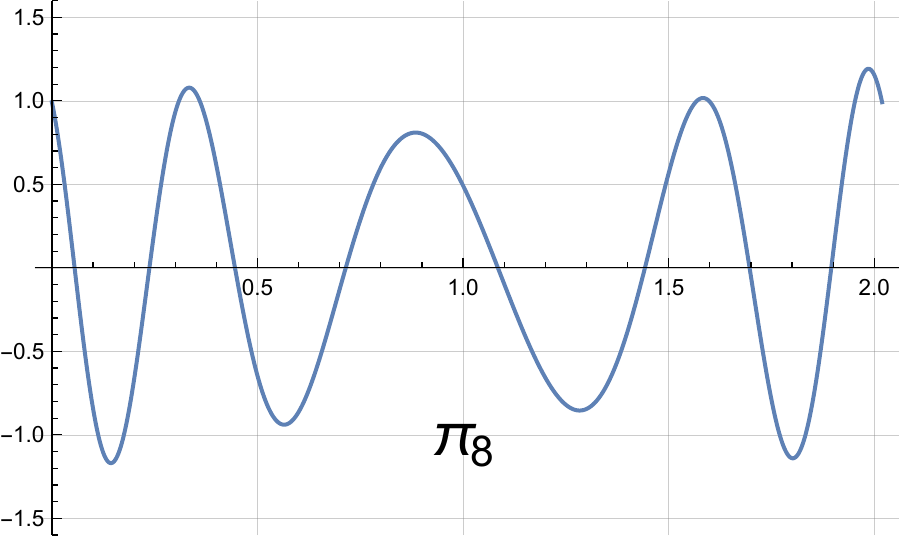}\\
\includegraphics[width=0.24\textwidth]{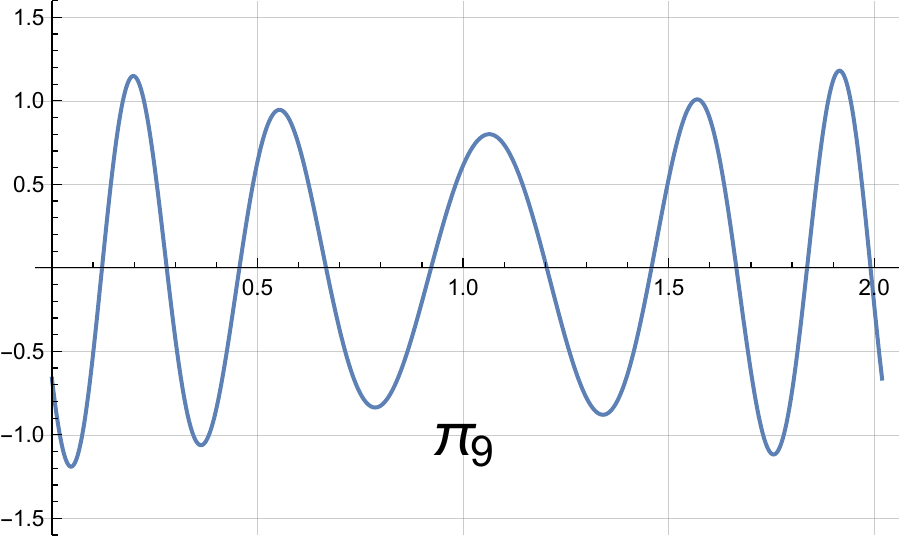}
\includegraphics[width=0.24\textwidth]{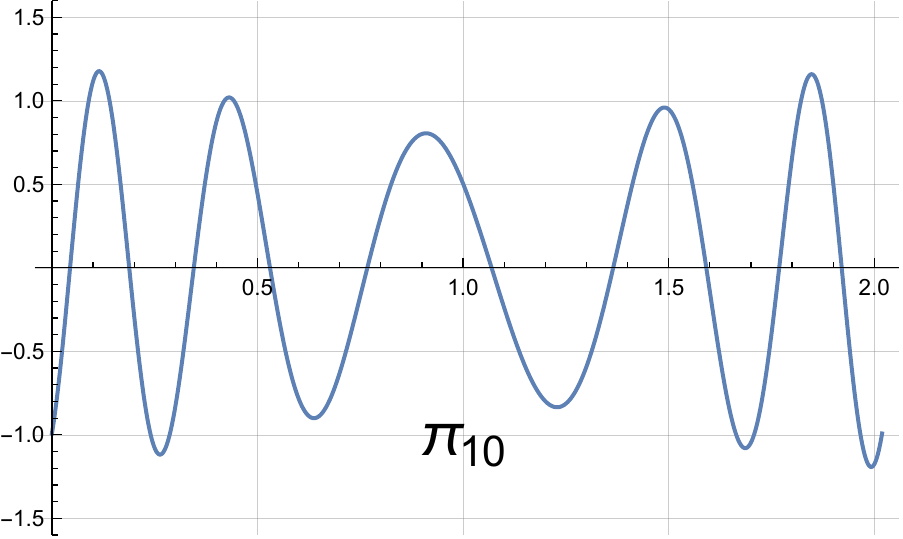}
\includegraphics[width=0.24\textwidth]{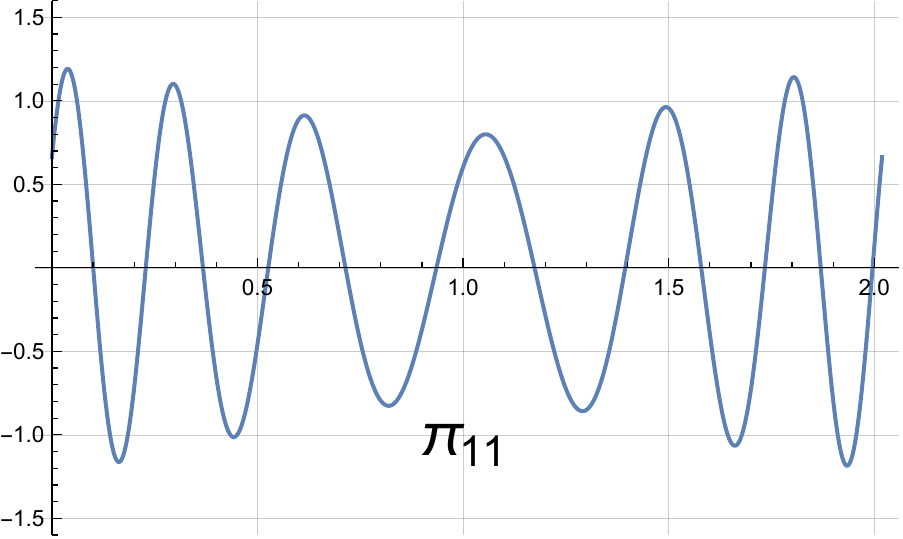}
\includegraphics[width=0.24\textwidth]{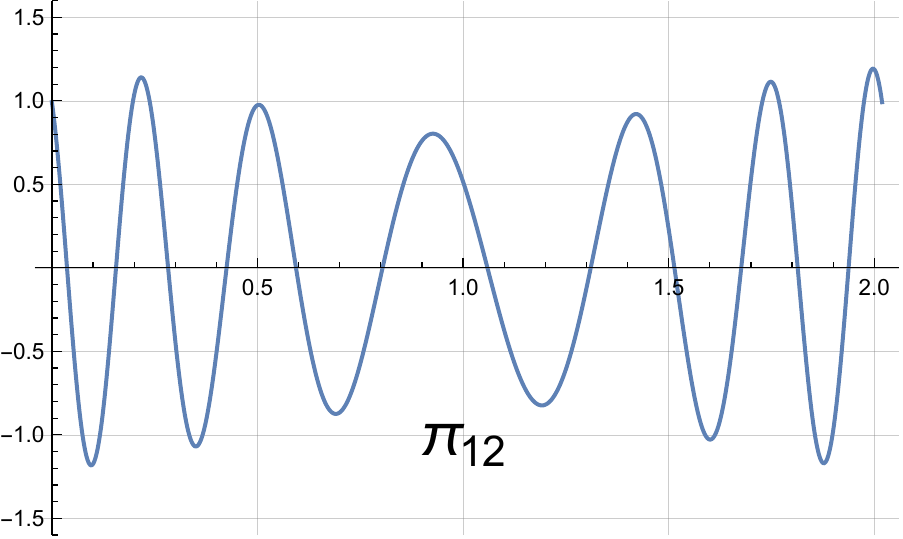}\\
\phantom{LA vistpa teresa}

\caption{The first few orthonormal sections; here $\pi _n\in \scr L(n\infty+\scr D)$. The elliptic curve is $Y^2=4X^3 - 28X+24 = 4(X-1)(X-2)(X+3)$. Here $\omega_1 \simeq 1.0094529, \ \ \omega_2 = 0.7422062\,i$. We have set $\scr D = \frac {\omega_1} 2\in \R$ and $\infty =0$. 
The contour $\gamma$ is  the segment $[\omega_2,\omega_2+2\omega_1]$ in $\C/2\omega_1\Z+2\omega_2\Z$; in the $X$--plane this is the segment $X\in [1,2]$ (on both sheets).
 Note that the  number of zeros of $\pi_n$ is not strictly monotonic with respect to the degree $n$. In particular some orthogonal sections have zeros outside of $\gamma$, differently from the case of orthogonal polynomials on the real line.
 }
 \label{Figure}
\end{figure}

A numerical evaluation can be performed. The resulting first few orthonormal sections are plotted in Fig. \ref{Figure} (with the independent variable $z=\omega_2+s$, and $s\in (0,2\omega_1)$. 
We observe that certain common theorems that apply to orthogonal polynomials do not apply here. In particular there are orthogonal sections of degree $n$ which have $n-2$ ``real'' (i.e. on $\gamma$) zeros. 

\paragraph{Some remarks.}
We conclude this section with some remarks. The author could not find any literature discussing orthogonal section of line bundles  in the sense of generalization of Pad\'e\ approximants with the partial exception of \cite{olver} where, however, only the orthogonal ``polynomials'' and not the Pad\'e\ problem are considered. Therefore there would be many questions regarding which of the classical results can be generalized in this context.

For example, (for the curves with antiholomorphic involution discussed in this last section) a natural question is where the zeros of the orthogonal sections are, and if something can be said for general (positive) measures. 
The ordinary proof of the reality and interlacing of orthogonal polynomials rely ultimately on the fact that polynomials are also an algebra, which is no longer the case in higher genus: indeed, the graded vector space $\bigoplus_{n\geq 0} \scr L(n\infty+\scr D)$ is the analog of the space of polynomials but is not an algebra. 

Even the simple example indicated above (genus $1$ and flat measure) would seem something of classical nature and possibly more properties of these orthogonal sections can be determined. 
For example it is tempting to conjecture that the number of zeros on the contour $\gamma$ (fixed by the anti-involution) should be increasing by $2g$ every $g$ steps and that an interlacing  of the zeros is still a universal feature. 

Much more interesting, and challenging, is the asymptotic description of the density of zeros, or even more, a strong asymptotic description of the orthogonal sections for large degree. 

In this context, one could hope to adapt the techniques of the Deift--Zhou Steepest descent for Riemann--Hilbert problems (for either fixed measures or scaling weights) as in the literature for orthogonal polynomials \cite{DKMVZ}. This indeed was the main impetus for seeking the  Riemann--Hilbert Problem \ref{RHP}. 
The immediate obstacle is the presence of the Tyurin data, which depend in a transcendental way on the measure.
\section{Construction of KP tau functions: generalizing algebro-geometric tau functions}
\label{KPsect}
A famous construction of Krichever's \cite{Krichever} gives rise to the so--called ``algebro--geometric'' solutions of the Kadomtsev--Petviashvili (KP) hierarchy. 
We are not recalling the whole construction here because it is well known in the community of integrable systems; for a modern review see \cite{BabelonBook}. Here we simply recall that the data are 
\begin{enumerate}
\item[-] a non-special divisor $\scr D$ of degree $g$, 
\item[-] 
a point $\infty\in \CC$ and a local coordinate $1/z(p)$ (such that $1/z(\infty)=0$).
\end{enumerate}
 This is  a subset of the data of our present setup: in addition to the above we have a measure $\d\mu$ on a contour $\g\subset \CC$. It is then natural to investigate if we can extend that construction. This is indeed possible as we see in Theorem \ref{mainKP}.

%\paragraph{Orthogonal polynomials: a brief reminder.}
%For the genus zero case we consider $\g=\R$ and $\d\mu$ a smooth measure. We suppose for ease of this reminder that the measure is compactly supported. 
%It can be shown that 
%\be
%\label{hankeltau}
%\tau_n(\bt ) := \int_{\R^n} \Delta^2(\vec x) \prod_{j=1}^n {\rm e}^{\xi(x_j;\bt)} \d\mu(x_j)
%\ee
%is a KP tau function (in the sense we recall below).  Here 
%\be
%\xi(x;\bt ) := \sum_{\ell\geq 1} t_\ell x^\ell,
%\ee
%where the series may be considered formal (or, less formally, one multiplies the measure $\d\mu$ by an arbitrary entire funtion ${\rm e}^\xi$).\\[2pt]
%

We are now going to consider the family of degree zero line--bundles $\scr L_\bt $ trivialized on the two sets of  a disk around $\infty$ and the punctured surface $\CC\setminus \{\infty\}$  with transition function ${\rm e}^{\xi(p;\bt)}$, where we have set 
\be
\xi(p;\bt):= \sum_{\ell\geq 1}t_\ell z(p)^\ell 
\ee 
for brevity. In concrete terms, a meromorphic section of $\scr L_\bt$ is simply a function $f(p)$ which is meromorphic on $\CC\setminus \{\infty\}$, with an essential singularity at $\infty$ and such that $f(p){\rm e}^{-\xi(p;\bt)}$ is meromorphic in a neighbourhood of $\infty$.

Then the symbol $\scr L_\bt(\scr D+n\infty)$ stands for the vector space of all functions $f(p)$ such that $f(p)$ has poles at $\scr D$ whose order does not exceed the multiplicity of the divisor and such that $f(p){\rm e}^{-\xi(p;\bt)} z(p)^{-n}$ is locally analytic near $\infty$. 

In Krichever's approach the Baker--Akhiezer function is a spanning element of $\scr L_\bt(\scr D)$ and in general one easily shows that 
\be
\dim_{\C} \scr L_\bt(\scr D+n\infty)\geq  n+1,
\ee
with the equality holding for a divisor $\scr D$ and $\bt$ in generic position. For convenience we denote with $\wh {\scr L}_\bt(\scr D)= \sum_{n\geq 0} \scr L_\bt(\scr D+n\infty)$. Namely, this is the infinite dimensional space of all meromorphic sections of $\scr L_\bt$ whose poles are at most at $\scr D$ and with order which does note exceed the multiplicity of the points of  $\scr D$. 

Consider now the following pairing on $\wh {\scr L}_\bt(\scr D)\otimes \wh {\scr L}_\bfs(\scr D)$:
\be
\label{pair}
\le\langle \ri\rangle_{\bt, \bfs} : \wh {\scr L}_\bt(\scr D)\otimes \wh {\scr L}_\bfs(\scr D)\to \C
\ee
given by 
\be
\le\langle \phi, \psi\ri\rangle_{\bt, \bfs} = \int_\gamma \phi(p) \psi(p) \d\mu(p)
\ee
Our ultimate goal is to define a sequence of  functions $\tau_n(\bt,\bfs),\  n\geq 0$ (see Def. \ref{defTau})  with the following properties:
\begin{enumerate}
\item $\tau_n(\bt,\bfs)=0$ if and only if the pairing \eqref{pair} restricted to  ${\scr L}_\bt(\scr D+(n-1)\infty)\otimes  {\scr L}_\bfs(\scr D+(n-1)\infty)$ is degenerate or $\dim_\C {\scr L}_{\bt}(\scr D)>1$ or $\dim_\C {\scr L}_{\bfs}(\scr D)>1$;
\item It satisfies the Kadomtsev--Petviashvili (KP) hierarchy in both sets of infinite variables $\bt, \bfs$;
\item It satisfies the $2$--Toda hierarchy.
\end{enumerate}
Before proceeding with this plan, we  provide a (formal) definition of KP tau functions  which is convenient for us; historically this is not the definition but a theorem that characterizes KP tau functions \cite{BabelonBook}. However it is expedient for us to flip history on its head and use this characterization as a definition.

\begin{definition}
\label{defHBI}
A (formal) KP tau function is a function $\tau (t_1,t_2,\dots) = \tau(\bt)$ of infinitely many variables that satisfies the {\em Hirota bilinear identity} (HBI) 
\be
\label{HBI}
\res{z=\infty}  \tau(\bt -[z^{-1}])\tau(\wt \bt + [z^{-1}]) {\rm e}^{\xi(z;\bt) - \xi(z;\wt \bt)} \d z  \equiv 0
\ee
for all $\bt,\wt \bt$. Here we use the notation 
\be
[z^{-1}] = \le(\frac 1 z, \frac 1 {2z^2}, \dots, \frac 1{\ell z^\ell}, \dots \ri).
\ee
\end{definition}
%{\color{orange}
%The proof that \eqref{hankeltau} satisfies the HBI \eqref{HBI} is provided in appendix \ref{} for ease of the reader: this statement is equivalent to stating that the partition function of a unitary ensemble of matrices is a KP tau function, which is known and can be proven in many different ways \cite{} (the proof we provide is yet different and hence of some interest).  More is actually true; one can view  $\tau_n$ as a function of the integer parameter $n$ as well, in which case it satisfies more bilinear identities that manifest the fact that it is a Toda tau function. We do not dwell on this issue in the higher genus case.}

\paragraph {\bf Notations for algebro-geometric objects.} 
We choose a Torelli marking $\{a_1,\dots, a_g, b_1,\dots, b_g\}$ for $\CC$ in terms of which we construct the classical Riemann Theta functions. We refer to \cite{Fay}, Ch. 1-2 for a review of these classical notions. Here we shall denote by $\Theta_\Delta$ the Theta function with characteristc $\Delta$, which is chosen as a nonsingular half-integer characteristics in the Jacobian of the curve. We remind the reader that this implies that $\Theta_\Delta(\mathbf z),\ \mathbf z\in \C^g$ is an odd function on $\mathbb J(\CC)$ and that the gradient at $\mathbf z=0$ does not vanish.
We also need the  normalized holomorphic differentials $\omega_1,\dots, \omega_g$ such that 
\be
\oint_{a_j} \omega_k = \delta_{jk}.
\ee
The Abel map $\u_\infty(p)$ will be defined with basepoint chosen at $\infty$:
\be
\u_\infty(p):= \int_\infty^p \le[{\omega_1\atop {\vdots\atop \ds\omega_g}}\ri]\label{Abel}.
\ee
Following a common use in the literature (see \cite{Fay}), we omit the notation of the Abel map when it is composed with the Riemann Theta function; to wit, for example, if $p\in \CC$ is a  point on the curve we shall write $\Theta(p)$ to mean $\Theta(\u_\infty(p))$. Similarly, if $\scr D= \sum k_j p_j$ is a divisor on $\CC$  with $k_j\in \Z$ and $p_j$ a collection of points, the writing $\Theta(\scr D)$ stands for $\Theta\le(\sum k_j \u_\infty(p_j)\ri)$. Finally we denote by $\K$ the vector of Riemann constants (e.g. \cite{Fay}, pag. 8). Note the $\K$ depends on the choice of basepoint of the Abel map.

Let $\Omega(p,q)$ be the ``fundamental bidifferential'' (\cite{Fay}, pag 20); 
\be
\label{defOmega}
\Omega(p,q) = \d_p \d_q \ln \Theta_\Delta(p-q).
\ee
This is the unique bi-differential with the properties that it is symmetric in the exchange of arguments, its $a$--periods vanish and it has a unique pole  for $p=q$ with bi-residue equal to one (more details can be found in loc. cit.). 

Let $\Omega_\ell,\  \ell\geq 1$ be the unique meromorphic differentials of the second kind with a single pole of order $\ell+1$ at the point $\infty$ and  such that 
\bea
\Omega_\ell(p) &= \le(\ell z(p)^{\ell-1} + \mathcal O(z(p)^{-2}) \ri)\d z(p), \ \ \ p\to \infty\nn\\
&\oint_{a_j} \Omega_\ell=0,\ \ \ j=1,\dots, g.
\eea
They can be written in terms of the fundamental bidifferential as follows
\be
\Omega_\ell(p) =- \res{q=\infty} z(q)^{\ell} \Omega(q,p).
\ee
With these notations we now remind that the (multi--valued) function $\Theta_\Delta(p)$ has $g$ zeros at the points $\infty$ and $\scr D_{\Delta}$ (a divisor of degree $g-1$). 
The following holomorphic differential:
\be
\omega_\Delta(p):= \sum_{\ell=1}^g \frac{\pa}{\pa u_\ell} \Theta_\Delta(\vec u)\bigg|_{\vec u=0} \omega_\ell(p)
\label{omegadelta}
\ee
vanishes at $2\scr D_\Delta$ (i.e. has only zeros of even order and double that of the points of $\scr D_\Delta$).
For later convenience we define the constant 
\be
\label{defCCC}
\CCC = -\lim_{p\to \infty} z(p) \Theta_\Delta(p).
\ee
From the definition of $\omega_\Delta$ \eqref{omegadelta} a simple local analysis shows that $\CCC$ can be also expressed as follows also the property
\be
\label{omeganorm}
\lim_{p\to\infty} z^2(p) \frac {\omega_{\Delta}(p)}{\d z(p)} =\CCC.
\ee

To shorten formulas we lift $\xi$ to a function in the neighbourhood of $\infty$ by using the local coordinate $z$:
\be
\xi(p;\bt):= \sum_{\ell\geq 1} t_\ell z(p)^\ell.\label{defxi}
\ee
Note that this (formal) function is defined only in the coordinate chart covered by our chosen local coordinate $z$. In terms of $\xi$ we define the  differential
\be
\label{deftheta}
\d\vartheta(p;\bt) := \sum_{\ell \geq 1} t_\ell \Omega_\ell(p) =- \res{q=\infty} \xi(q;\bt) \Omega(q,p).
\ee
This can be described as the unique differential on $\CC\setminus \{\infty\}$ with prescribed singular part near $\infty$ given by $\d \xi(p;\bt)$ and normalized so that its $a$--periods vanish.
We denote by $\vartheta(p;\bt) $  its antiderivative with the constant of integration adjusted so that 
\be
\vartheta(p;\bt) - \xi(z(p);\bt) = \mathcal O(z(p)^{-1}), \ \ p\to \infty.
\ee
I.e., $\vartheta(p;\bt) = \sum_{\ell\geq 1} t_\ell z(p)^\ell + \mathcal O(z(p)^{-1}).$

Finally we denote by $\mb V(\bt)\in \C^g$ the vector in the Jacobian $\mb J(\CC)$ with components
\be
\label{defV}
\mb V_j(\bt):=\frac 1{2i\pi} \oint_{b_j} \d \vartheta(p;\bt) = -\sum_{\ell\geq 1} t_\ell \res{p=\infty} z(p)^{\ell}\omega_j(p).
\ee 
The second equality is a consequence of Riemann bilinear identities.
For $x\in \CC$ a point of the Riemann surface in the coordinate patch of $z$, we use the notation
\be
\label{shortmiwa}
[x]:= [z(x)^{-1}] = \le(\frac 1{z(x)}, \frac 1 {2z(x)^2} ,\dots, \frac 1{n z(x)^n}, \dots,\ri).
\ee

The role of the Cauchy kernel \eqref{Cauchy} is now played by the following generalization.

\begin{definition}
The twisted Cauchy kernel is the following expression:
\be
\label{twistCauchy}
\Cauchy(p,q;\bt) = {\rm e}^{\vartheta(q;\bt)-\vartheta(p;\bt )} \frac {\Theta( p-q- \mb F(\bt)) \Theta(p-\scr D -\K)\Theta_\Delta(q)\omega_\Delta(p)}{\Theta(\mb F(\bt))\Theta_\Delta(p-q) \Theta_\Delta(p)\Theta(q-\scr D -\K)},
\ee
where 
\be
\mb F(\bt) := \mb V(\bt) - \u_\infty(\scr D)-\K.
\label{defF}
\ee
\end{definition}
It can be characterized as the unique kernel on $\CC\setminus \{\infty\}$ which is a differential w.r.t. $p$,  meromorphic function w.r.t. $q$  and with the properties:
\begin{enumerate}
\item w.r.t. $p$ it has  zero divisor  $\geq \scr D$ and a simple pole at $q$ of residue $1$;
\item w.r.t. $q$ it has pole divisor $\geq -\scr D$ and a simple pole at $p$;
\item when $p,q$ are in a neighbourhood of $\infty$ and hence fall within the same coordinate patch $z$, it can be written 
\be
\Cauchy(p,q;\bt) = {\rm e}^{\xi(q;\bt)-\xi(p;\bt)} \le(\frac 1{z-w} + \mathcal O(z^{-2})\mathcal O(w^{-1}) \ri) \d z
\ee
where $z = z(p)$ and $w = z(q)$.\footnote{We hope that the notation here is not too confusing; $z$ is the value $z(p)$ and $w$ is the value of $z(q)$. They are simply the local coordinates of the points $p,q$ in the coordinate $z$ near $\infty$.}
\end{enumerate}
We observe that for $\bt =0$ it coincides with the Cauchy kernel \eqref{Cauchy}.
\begin{remark}
\rm 
Observe that the Cauchy kernel ceases to exist when $\Theta(\mb F(\bt))=0$; this corresponds, by a consequence of Riemann vanishing theorem and Riemann--Roch's theorem to the statement that $\dim_\C {\scr L }_{\bt}(\scr D)> 1$, namely, that there is no Baker-Akhiezer function in Krichever's setup. \hfill $\triangle$
\end{remark}

The twisted kernel \eqref{twistCauchy} allows us to construct spanning sections of $\scr L_\bt(\scr D+n\infty)$ by the following formula 
\be
\zeta_n(q;\bt):= \res{p=\infty} z(p)^n {\rm e}^{\xi(p;\bt)} \Cauchy(p,q;\bt).
\ee
A simple local analysis shows that the behaviour of $\zeta_n$ near $\infty$ is of the form 
\be
\label{zetainfty}
\zeta_n(x;\bt) = {\rm e}^{\xi(x;\bt)} \bigg(z(x)^{n} + \mathcal O(z(x)^{-1}) \bigg), 
\ee 
and hence $\zeta_n(q;\bt)\in \scr L_\bt(\scr D+n\infty)\setminus  \scr L_\bt(\scr D+(n-1)\infty)$ so that  $ \scr L_\bt(\scr D+n\infty) = {\rm Span}_{\C}\le\{\zeta_a; \ \ a=0,\dots, n\ri\}$.
\begin{remark}
Observe that $\zeta_0(p;\bt)$ is the usual Baker--Akhiezer function of Krichever's; then another convenient spanning set can be defined by $\pa_{t_1}^\ell \zeta_0(p;\bt)$, $\ell=0,\dots, n$. 
\end{remark}

\paragraph{Biorthogonal sections.}
A simple exercise shows that the following two sections  $P_n\in \scr L_\bt(\scr D+n\infty)$ and $Q_n\in \scr L_\bfs(\scr D+n\infty)$ are ``biortogonal'' with respect to the pairing \eqref{pair} in the sense that $P_n\perp \scr L_\bt(\scr D+(n-1)\infty)$ and $Q_n\perp \scr L_\bt(\scr D+(n-1)\infty)$:
\bea
\label{defPn}
P_n(x;\bt,\bfs)&:= \det \le[\begin{array}{cccc}
\mu_{00} & \dots &\mu_{n,0}\\
\vdots &&\vdots\\
\mu_{n-1,0} & \dots & \mu_{n-1,n}\\
\zeta_0(x;\bt) & \dots & \zeta_n(x;\bt)
\end{array}\ri]
\ ,\cr
Q_n(x;\bt,\bfs)&:= \det \le[\begin{array}{ccc|c}
\mu_{00} & \dots &\mu_{n-1,0} &\zeta_0(x;\bfs)\\
\vdots &&&\vdots\\
\mu_{n-1,0} & \dots & \mu_{n-1,n}\\
\mu_{n0} & \dots & \mu_{n,n-1} &\zeta_n(x;\bfs)
\end{array}\ri],
\eea
where we have introduced the generalized bi-moments
\be
\label{defmu}
\mu_{ab} = \mu_{ab}(\bt,\bfs) := \le\langle \zeta_a(\bullet ;\bt), \zeta_b(\bullet; \bfs)\ri\rangle= \int_\gamma\zeta_{a}(p;\bt) \zeta_b(p;\bfs)  \d\mu(p) .
\ee
For example
\be
\label{biorthogonality}
\int_\g P_n(x;\bt, \bfs) \zeta_a(x;\bfs)\d\mu(x)=0  \ \ \forall a=0,\dots, n-1.
\ee

We now come to the main object of the section;
\begin{definition}[The Tau function]
\label{defTau}
The Tau function is defined by 
\bea
\label{tau}
\tau_n(\bt ,\bfs):=
& \frac 1{n!}\Theta(\mb F(\bt) )\Theta(\mb F(\bfs) ) {\rm e}^{Q(\bt) + Q( \bfs) + nA(\bt) + nA(\bfs)} 
\times\cr
&\times  \int_{\gamma^n} \det\big[\zeta_{a-1}(r_b;\bt)\big]_{a,b=1}^n\det\big[\zeta_{a-1}(r_b;\bfs)\big]_{a,b=1}^n  \prod_{j=1}^n \d\mu(r_j)=\\
=& \Theta(\mb F(\bt) )\Theta(\mb F(\bfs) ) {\rm e}^{Q(\bt) + Q( \bfs) +nA(\bt)+nA(\bfs)} \det \bigg[\mu_{ab}(\bt,\bfs) \bigg]_{a,b=0}^{n-1} \label{tau2}
\eea
The expression  $Q(\bt)$ in \eqref{tau} is the quadratic form 
\be
\label{defQ}
Q(\bt) := \frac 1 2 \res{p=\infty}\res{q=\infty} \xi(p;\bt) \xi(q;\bt) \Omega(p,q) .
%=\sum_{\ell,m\geq 1} t_\ell t_m Q_{\ell,m}\ ,\ \ \ Q_{\ell,m} = \res{p=\infty}z(p)^m \Omega_{\ell}(p)= Q_{m,\ell}.
\ee
and $A (\bt)$ is the linear form 
\be
\label{An}
A(\bt) = \sum_{\ell\geq 1} \ell t_\ell c_\ell =-\res{p=\infty} \d \xi(p;\bt) \ln\bigg( z(p)\Theta_\Delta(p) \bigg)
\ee
where $c_\ell$ are the coefficients of the expansion of $\ln (\Theta_\Delta(x)z(x))$ near $\infty$ in the coordinate $z(x)$:
\be
\ln (\Theta_\Delta(x)z(x)) = \sum_{\ell\geq 0} \frac {c_\ell}{z(x)^\ell}.
\ee
\end{definition}
The equivalence of \eqref{tau} and \eqref{tau2} follows from the Andreief identity \cite{Andreief}.
The formula \eqref{tau2} shows clearly that $\tau_n(\bt,\bfs)=0$ if and only if the pairing \eqref{pair}  is degenerate or $\Theta(\mb F(\bt))\Theta(\mb F(\bfs))=0$. 
\begin{remark}
The expression $\Theta(\mb F){\rm e}^{Q(\bt)}$ is the algebro--geometric KP tau function corresponding to Krichever's construction (see \cite{BabelonBook}, Ch. 8). Thus, for $n=0$ the tau function is just the product of two independent Krichever algebro-geometric KP tau functions.
For $n\geq 1$ the determinant of the moments ``entangles'' them into a single object.
\end{remark}

We now state the first main theorem 
\bet
\label{mainKP}
For every $n\in \mathbb N$, the tau function $\tau_n (\bt,\bfs)$ is a KP tau function separately in each set of variables $\bt, \bfs$. Namely it satisfies the two Hirota Bilinear Identities
\bea
\label{HBIt}
\res{x=\infty}  \tau_n(\bt -[x], \bfs)\tau_n(\wt \bt + [x], \bfs) {\rm e}^{\xi(x;\bt) - \xi(x;\wt \bt)} \d z(x)  \equiv 0
\\
\label{HBIs}
\res{x=\infty}  \tau_n(\bt , \bfs-[x])\tau_n(\bt, \wt \bfs + [x]) {\rm e}^{\xi(x;\bfs) - \xi(x;\wt \bfs)} \d z(x)  \equiv 0
\eea 
where $[x]$ is defined in \eqref{shortmiwa}  and $\xi$ is defined in \eqref{defxi}.
\eet
It is clear that the roles of $\bt, \bfs$ are completely symmetric in the definition \eqref{tau}, and hence it suffices to give the proof of \eqref{HBIt}. For this reason we will focus on the $\bt$ dependence, leaving the reader the exercise to reformulate similar statements for the $\bfs$ dependence. 

The argument in the residue formula \eqref{HBIt} is usually split into the product of the so--called Baker--Akhiezer (and dual partner) functions. In fact these functions have their own definition (see for example \cite{SegalWilson}) and their relationship with the tau function is rather a theorem that generally goes under the name  of {\it Sato's formula}. Here we do not make this distinction because it is not relevant to the paper and we identify the Baker--Akhiezer functions with their expression in terms of Sato's formula. 

\bp
\label{propBaker}
The Baker--Akhiezer function is 
\be
\label{B}
\frac{\tau_n(\bt - [x];\bfs)}{\tau_n(\bt,\bfs)} {\rm e}^{\xi(x;\bt)} =   \frac{P_n(x;\bt, \bfs)}{\det[\mu_{ab} (\bt,\bfs) ]_{a,b=0}^{n-1}z(x)^n} \frac {-\CCC} {\Theta_\Delta(x)} \sqrt{\frac{ \omega_\Delta(x)}{\CCC\d z(x)}}
\ee
where $P_n$ is the biorthogonal section defined by  \eqref{defPn} (the constant $\CCC$ is defined in \eqref{defCCC}).
\ep 
The proof is in Section \ref{proofpropBaker}. The second component of the HBI's is the dual Baker function. For this reason we need the analog of Prop. \ref{propBaker} with the opposite shift in the times.
\bp
\label{propdualBaker}
The dual Baker function is 
\be
\label{dualB}
\frac{\tau_n(\bt + [x];\bfs)}{\tau_n(\bt;\bfs)} {\rm e}^{ -\xi(x;\bt)} = 
\frac{-z(x)^{n} \Theta_\Delta(x)}{\CCC \det\big[\mu_{ab}(\bt,\bfs)\big]_{a,b=0}^{n-1}}\sqrt{\frac{\CCC }{\omega_\Delta(x) \d z(x)}}\mathfrak R_n(x;\bt,\bfs)
\ee
where $\mathfrak R_n(x;\bt,\bfs)$ is the following differential with a discontinuity across $\gamma$:
\be
\mathfrak R_n(x;\bt,\bfs):=  \int_{r\in\gamma} \Cauchy(x,r;\bt) Q_{n-1}(r;\bt,\bfs) \d \mu(r)
\ee
and $Q_n$ is the biorthogonal section \eqref{defPn}. 
The jump discontinuity of $\mathfrak R_n$ across the contour $\gamma$ is given by 
\be
\label{Rjump}
\mathfrak R_n(x;\bt,\bfs)_+-\mathfrak R_n(x;\bt,\bfs)_-=  2i\pi Q_{n-1}(x;\bt,\bfs), \ \ \ x\in \gamma.
\ee
\ep 
The proof is in Section \ref{proofpropdualBaker}.\\[15pt]

With the aid of the  two Propositions \ref{propBaker}, \ref{propdualBaker} the proof of the main theorem is now a simple conclusion.

\noindent {\bf Proof of Theorem \ref{mainKP}.} Using \ref{dualB} and \ref{B} for the tau functions we see that their product in \eqref{HBIt} extends to a well--defined {\it holomorphic} differential in the variable $x$ defined on $\CC \setminus \gamma\cup\{\infty\}$. Thus we need to compute the following residue:
\bea
\label{res}
\res{x=\infty} P_n(x;\bt, \bfs) \mathfrak R_n(x;\wt \bt, \bfs) .
\eea
The differential $\mathfrak R_n$ has a jump discontinuity across the contour $\gamma$, an essential singularity at $\infty$ and it is otherwise holomorphic with zeros at $\scr D$ that cancel the poles of $P_n$.

Thus, the computation of the residue \ref{res} can be performed alternatively (Cauchy's theorem) by integrating along the contour $\gamma$ the jump discontinuity of the integrand using \eqref{Rjump}  and hence  
\bea
\res{x=\infty} P_n(x;\bt, \bfs) \mathfrak R_n(x;\wt \bt, \bfs)=\int_\gamma P_n(x;\bt,\bfs) Q_{n-1}(x;\wt \bt, \bfs)\d\mu(x)
\eea
Since $ Q_{n-1}(x;\wt \bt, \bfs)\in \scr L_{\bfs}(\scr D+(n-1)\infty)$, it follows that the integral vanishes because of the orthogonality of $P_n$ to the whole subspace \eqref{biorthogonality}.
\QED

\subsection{The $2$--Toda hierarchy}

A simple modification of the computation above allows us to prove the following corollary, which gives a modification of the 2-Toda bilinear identities \cite{AdlerVanMoerbeke, UenoTakasaki}.
\begin{corollary} 
The following modified $2$--Toda bilinear identities hold:
\bea
&\res{x=\infty}{\tau_n(\bt-[x]; \bfs)\tau_{m+1} (\wt \bt+[x]; \wt \bfs)} 
\frac{{\rm e}^{\xi(x;\bt)-\xi(x;\wt\bt)+A(\wt \bt - \bt)}\d z(x)}{z(x)^{m-n+1}}
=\cr
&= \res{x=\infty} {\tau_{n+1}(\bt; \bfs+[x])\tau_{m} (\wt \bt; \wt \bfs-[x])}\frac{{\rm e}^{\xi(x;\wt \bfs)-\xi(x;\bfs) + A(\bfs - \wt \bfs)}\d z(x)}{z(x)^{n-m+1}}\label{2todaHBI}
\eea
where $A$ it the linear expression \eqref{An} in terms of the times $\bt$.
\end{corollary}
{\bf Proof.}
For brevity we denote the pre-factor in the Definition \ref{defTau} of the tau function  (formula  \eqref{tau}) by 
\be
W_n(\bt, \bfs):= {\rm e}^{Q(\bt)+Q(\bfs)+nA(\bt)+nA(\bfs)} \Theta(\mb F(\bt))\Theta(\mb F(\bfs))
\label{defWn}
\ee
We can recast  \eqref{B} \eqref{dualB} as
\bea
\tau_n(\bt-[x]; \bfs) {\rm e}^{\xi(x;\bt)}&= W_n(\bt, \bfs) \frac{P_n(x;\bt, \bfs)}{z(x)^n} \frac {1} {\Theta_\Delta(x)} \sqrt{\frac{\CCC \omega_\Delta(x)}{\d z(x)}}
 \\
\tau_n(\bt+[x]; \bfs) {\rm e}^{-\xi(x;\bt)}&= W_n(\bt, \bfs) \frac{z(x)^{n} \Theta_\Delta(x)}{-\CCC }\sqrt{\frac{\CCC }{\omega_\Delta(x) \d z(x)}}\mathfrak R_n(x;\bt,\bfs).
 \eea
Thus we have 
\bea
{\tau_n(\bt-[x]; \bfs)\tau_{m+1} (\wt \bt+[x]; \wt \bfs)}\frac{\d z(x)}{z(x)^{m+1-n}} = \cr
=W_n(\bt, \bfs) W_{m+1}(\wt \bt, \wt \bfs)
 \frac{P_n(x;\bt, \bfs) \mathfrak R_{m+1}(x;\wt\bt,\wt \bfs)} {\CCC } 
\eea
Then after taking the residue and converting the residue to an integral over $\gamma$ as in Theorem \ref{mainKP}, we obtain 
\be
\res{x=\infty}{\tau_n(\bt-[x]; \bfs)\tau_{m+1} (\wt \bt+[x]; \wt \bfs)}\frac{{\rm e}^{\xi(x;\bt)-\xi(x;\wt\bt)} \d z(x)}{z(x)^{m+1-n}} =
W_n(\bt, \bfs) W_{m+1}(\wt \bt, \wt \bfs) \int_{\gamma} P_n(x;\bt,\bfs) Q_{m}(x;\wt \bt,\wt \bfs)\d\mu(x).
\label{347}
\ee
Repeating the same computation on the right side of \eqref{2todaHBI}, we have to use the formul\ae\ (which are simply a rephrasing of Prop. \ref{propBaker} and Prop. \ref{propdualBaker})
\bea
\tau_n(\bt; \bfs-[x]) {\rm e}^{\xi(x;\bfs)}&=
W_n(\bt, \bfs)  \frac{Q_n(x;\bt, \bfs)}{z(x)^n} \frac {1} {\Theta_\Delta(x)} \sqrt{\frac{\CCC  \omega_\Delta(x)}{\d z(x)}}
 \\
\tau_n(\bt; \bfs+[x] ) {\rm e}^{-\xi(x;\bfs)}&= 
W_n(\bt, \bfs) 
\frac{z(x)^{n} \Theta_\Delta(x)}{-\CCC}\sqrt{\frac{\CCC }{\omega_\Delta(x) \d z(x)}}\mathfrak S_n(x;\bt,\bfs)\\
&\mathfrak S_n(x;\bt,\bfs)=\int_\gamma \Cauchy(x,r;\bfs)P_{n-1}(r;\bt,\bfs)\d\mu(r).
 \eea
Using these formulas on the right side of \eqref{2todaHBI}  we obtain
\bea
\res{x=\infty} {\tau_{n+1}(\bt; \bfs+[x])\tau_{m} (\wt \bt; \wt \bfs-[x])}\frac{{\rm e}^{\xi(x;\bfs)-\xi(x;\wt\bfs)}\d z(x)}{z(x)^{n-m+1}}=\cr =
W_{n+1}(\bt, \bfs) W_{m}(\wt \bt, \wt \bfs)
\int_{\gamma} P_n(x;\bt,\bfs) Q_{m}(x;\wt \bt,\wt \bfs)\d\mu(x)
\eea
where we observe that the integral  is the same as \eqref{347}.
Now, the ratio of the constants gives
\be
\frac{W_n(\bt, \bfs) W_{m+1}(\wt \bt, \wt \bfs)}{
W_{n+1}(\bt, \bfs) W_{m}(\wt \bt, \wt \bfs)
}= {\rm e}^{A(\wt \bt)-A(\bt) + A(\wt \bfs)-A(\bfs) },
\ee
and this produces the statement of the theorem.
\QED

We note that the bilinear identities \eqref{2todaHBI} reduce to the standard ones in \cite{AdlerVanMoerbeke} in genus $g=0$ where the linear form $A$ vanishes.

\paragraph*{Acknowledgements.}
The work  was supported in part by the Natural Sciences and Engineering Research Council of Canada (NSERC) grant RGPIN-2016-06660.

\appendix
\renewcommand{\theequation}{\Alph{section}.\arabic{equation}}

\section{Proofs}
\subsection{The Sato shift}
Here we call ``Sato shift'' the shift of times $\bt$ occurring in Sato's formulas for the Baker--Akhiezer functions (Prop. \ref{propBaker}, \ref{propdualBaker}). 
The following lemmas show how the various ingredients of our formula transform when $\bt \mapsto \bt \pm [x]$ (with the definition \eqref{shortmiwa}). This is all in preparation of expressing the tau function and the HBI \eqref{HBI}.
These lemmas can be traced in the literature in several places but we refer comprehensively (at least for part of them) to \cite{BabelonBook}. We provide our own proofs for convenience of the reader.

The simplest result is the following one, which is a simple exercise from the definition \eqref{An}):
\be
\label{shiftAn} 
{\rm e}^{A(\bt \pm [x])}= {\rm e}^{A(\bt)}\le( \frac {\Theta_\Delta(x) z(x)}{-\CCC }\ri)^{\pm 1}\ .
\ee
We remind the reader of the definition of $\varkappa$ in \eqref{defCCC} and  of our stipulation (discussed after \eqref{Abel})  that the writing $\Theta_\Delta(x)$ is a shorthand for $\Theta_\Delta(\int_\infty^x \vec \omega)$.
\begin{lemma}
\label{lemmaV}
Under the Sato shift, the  vector $\mb V(\bt)$ defined in \eqref{defV} transforms as follows 
\be
\label{shiftV}
\mb V(\bt\pm [x]) = \mb V(\bt) \mp \int_\infty^x \vec \omega.
\ee
\end{lemma}
{\bf Proof.}
Observe that $\xi(p;\bt-[x]) =\xi(p;\bt)-  \sum_{\ell\geq 1} \frac 1 \ell  \le(\frac {z(p)}{z(x)}\ri)^\ell = \xi(p;\bt) + \ln \le(1 - \frac {z(p)}{z(x)} \ri)$, where the resummation holds as long as $|z(x)|>|z(p)|$. Using this simple observation we obtain, using \eqref{defV}:
\bea
\mb V_\ell(\bt -[x]) - \mb V_\ell(\bt)
& =\frac 1{(2i\pi)^2}\oint_{|z(q)|=R} \ln\le(1- \frac{ z(q)}{z(x)} \ri) \oint_{p\in b_\ell}\Omega(q,p)=\cr
&= 
\frac 1{2i\pi}\oint_{|z(q)|=R} \ln\le(1- \frac{ z(q)}{z(x)} \ri) \omega_\ell(q)
=\cr
&=-\frac 1{2i\pi} \oint_{|z(q)|=R}    \frac {\d z(q)}{z(q)-z(x)} \u(q) = \u(x)
\eea
where the contour integral is {\it counterclockwise}\footnote{Since the coordinate local coordinate at $\infty$ is $\frac 1{z(p)}$ the integral defining $(-\res{p=\infty})$ in the $z$--plane is a counterclockwise large circle.} in the $z(q)$--plane and $|z(x)|>R$.
\QED

 The quadratic form \eqref{defQ} is well known in the Krichever approach \cite{BabelonBook}. 
The main property that we are going to use is reported below
\bp
\label{propQ}
The quadratic form \eqref{defQ} has the properties:
\bea
\label{Qshift+}
{\rm e}^{Q(\bt - [x])} ={\rm e}^ {Q(\bt) + \vartheta(x;\bt)-\xi(z(x);\bt)} \frac {\CCC }{ \Theta_\Delta(x)}\sqrt{ \frac{\omega_\Delta(x)}{\CCC \d z(x)}}
\\
\label{Qshift-}
{\rm e}^{Q(\bt + [x])} ={\rm e}^ {Q(\bt) -\vartheta(x;\bt)+\xi(z(x);\bt)} \frac {\CCC }{ \Theta_\Delta(x)}\sqrt{ \frac{\omega_\Delta(x)}{\CCC \d z(x)}}
\eea
with the notation \eqref{shortmiwa} and $\CCC$ the constant defined in \eqref{defCCC}.
\ep 
{\bf  Proof.}
\label{proofpropQ} 
From the definition of $\xi$ \eqref{defxi} and  \eqref{shortmiwa} it follows that $\xi(q;\bt\pm [x]) = \xi(q;\bt) \mp \ln (1 - z(q)/z(x))$. This computation assumes that $|z(x)|>|z(q)|$ and hence, to make rigorous sense, we should realize the residues in \eqref{defQ} as counterclockwise  contour  integrals in the $z$--plane along  the circle $|z(q)|=R$ (with $R<|z(x)|$). Keeping this in mind, denote temporarily by $\mathcal Q(\bt, \bt)$ the polarization of the quadratic form $Q$: then we need to compute
\be
Q(\bt \mp [x]) = Q(\bt) \mp 2 \mathcal Q(\bt, [x]) + Q([x]).\label{expan}
\ee
We start from the last term $Q([x])$ and  we compute it using local coordinates $z = z(q), \wt z= z(p)$, and $w=z(x)$ letting $F(z,\wt z) = \Theta_\Delta(q-p)$.  We then  have, using integration by parts and the Cauchy theorem:
\bea
2Q([x])= \oint_{|z|=R} \frac {\d z}{2i\pi}\oint_{|\wt z|= R+\epsilon} \frac {\d \wt z}{2i\pi} \ln \le(1- \frac z w \ri) \ln \le(1- \frac {\wt z} w \ri) \frac {\d^2}{\d z \d \wt z} \ln F(z, \wt z) = \cr
%\cr
%=-\oint_{|z|=R} \oint_{|\wt z|= R+\epsilon}  \frac {\d \wt z} {(\wt z -w)2i\pi}  \ln \le(1- \frac { z} w \ri) \frac {\d}{ \d z} \ln \le(\frac{F(z, \wt z) }{F(z,\infty)}\ri)\frac {\d  z}{2i\pi}=\cr
=  \oint_{| z|= R}  \ln \le(1- \frac { z} w \ri) \frac {\d}{ \d  z} \ln \le(\frac{F(z, w)}{F(z,\infty)}\ri) \frac{\d z}{2i\pi}
\eea
 The logarithm now has a branch-cut from $z=w$ to $ z=\infty$ and is analytic along $|z|=R$ (recall that $|w|>R$) so that  can integrate by part along the circle obtaining
\bea
\label{415}
2Q([x])= -\oint_{| z|= R}\ln \le(\frac{F(  z,w)}{F(z,\infty)}\ri) \frac {\d  z}{( z-w)2i\pi} .
\eea
The computation of this last integral needs to be done with care because  of the   branch-cut.
 We   regularize the integral by   adding $\oint_{|z|=R} \ln (\frac {1}{w-z})\frac{\d z}{ z-w}$, which is zero because it is analytic in $|z|<R$ (since $|w|>R$: the branchcut of $\ln$ is from $z=w$ to $z=\infty$).  Then \eqref{415} becomes
\bea
2Q([x])=-\oint_{| z|= R+\epsilon}\frac {\ln \frac{F(z,w)}{(z-w) F(z,\infty) }\d z}{( z-w)2i\pi}.
\eea
The integrand is now meromorphic for $|z|>R$ with only first order poles at $z=w$ and $z=\infty$ and hence we obtain 
\be
2Q([x])=\ln\le( \frac{\pa_z F(z,w)\bigg|_{z=w}}{F(w,\infty)} \ri) - \ln \frac{ F(w,\infty)}{-\CCC } 
\ee
where $\CCC  = -\lim_{z\to \infty} z{F(z,\infty)}$ (recall that $F(z(p),\infty)= \Theta_\Delta(\int_{\infty}^p \vec \omega)$ vanishes of first order as $p\to\infty$).
The derivative of $F(z,w)$ on the diagonal in the local coordinate $z$ is precisely $\omega_\Delta(p)/\d z(p)$ and hence we have the final result 
\be
{\rm e}^{2Q([x])} = \frac {{\CCC }\omega_{\Delta}(x)}{\Theta_\Delta(x)^2\d z(x)}.
\ee
Here  the constant guarantees that the right side tends to one as $x\to\infty$ as  the left side does.
To complete the proof we need to evaluate $2\mathcal Q(\bt, [x])$ in \eqref{expan}; using the definition \eqref{defQ} we have
\be
\mp 2\mathcal Q(\bt, [x])= \frac{\pm 1}{(2i\pi)^2}  \oint_{|z(p)|=\wt R} \oint_{|z(q)|=R} \xi(p;\bt) \ln\le(1- \frac {z(q)}{z(x)} \ri) \Omega(p,q),
\ee 
with $R, \wt R<|z(x)|$.
A simple computation following the same steps as above yields then the regular part near $x=\infty$ of the Abelian integral $\vartheta(x;\bt)$, namely,
\be
\mp 2\mathcal Q(\bt, [x])=\mp\xi(x;\bt)\pm\vartheta(x;\bt),
\ee 
which completes the proof.
%
%
%\bea
%2\mathcal Q(\bt, [x]) = -\frac 1{(2i\pi)^2} \oint_{p} \oint_q \xi(p;\bt) \ln\le(1- \frac {z(q)}{z(x)} \ri) \Omega(p,q)=\\
%=\frac 1{(2i\pi)^2} \oint_{p} \oint_q\xi(p;\bt) \frac {\d z(q)}{z(q)-z(x)}  \d_p\ln \le(\frac{\Theta_\Delta(p-q)}{\Theta_\Delta(q)}\ri)
%=\\
%=-\frac 1{2i\pi} \oint_{p} \xi(p;\bt)   \d_p\ln \le({\Theta_\Delta(p-x)}\ri)
%\eea
%Since $|z(x)|>|z(p)|$ (the integral is counterclokwise in $z(p)$--plane) that integral is equal to 
%\bea
%2\mathcal Q(\bt, [x]) =-\frac 1{2i\pi} \oint_{|z(p)|<|z(x)|} \xi(p;\bt)   \d_p\ln \le({\Theta_\Delta(p-x)}\ri)
%=\\
%=\xi(x;\bt) -\frac 1{2i\pi} \oint_{|z(p)|>|z(x)|} \xi(p;\bt)   \d_p\ln \le({\Theta_\Delta(p-x)}\ri) 
%=\\ 
%=\xi(x;\bt)-\vartheta(x;\bt).
%\eea
\QED
We will also need the formula for the Sato shift on the Abelian integral $\vartheta$:
\begin{lemma}
\label{lemmaGshift}
We have the formula; 
\be
\label{Gshift}
{\rm e}^{\vartheta(p;\bt\pm [x])} = {\rm e}^{\vartheta(p;\bt\pm [x])} \le(\frac {\Theta_\Delta(p-x)}{\Theta_\Delta(p)} \ri)^{\mp 1}.
\ee
\end{lemma}
{\bf Proof.}
Observe that $\xi(p;\bt-[x]) =\xi(p;\bt)-  \sum_{\ell\geq 1} \frac 1 \ell  \le(\frac {z(p)}{z(x)}\ri)^\ell = \xi(p;\bt) + \ln \le(1 - \frac {z(p)}{z(x)} \ri)$, where the resummation holds as long as $|z(x)|>|z(p)|$.  Moreover, since $\Omega(p,q)$ is given by \eqref{defOmega} we can use integration by parts in the computation below: note that $-\res{q=\infty}$ is an integration in the counterclockwise orientation in the $z(q)$--plane.
Using these  observations we obtain:
\bea
\d \vartheta(p;\bt -[x]) - \d\vartheta(p;\bt)
\mathop{=}^{\eqref{deftheta}}
\frac 1{2i\pi}\oint_{|z(q)|=R} \ln\le(1- \frac{ z(q)}{z(x)} \ri)\d_p\d_q\ln \le(\frac{\Theta_\Delta(p-q)}{\Theta_{\Delta}(p)} \ri)=
\cr
=-\frac 1{2i\pi} \oint_{|z(q)|=R}    \frac {\d z(q)}{z(q)-z(x)}\d_p\ln \le(\frac{\Theta_\Delta(p-q)}{\Theta_{\Delta}(p)} \ri) = \d_p \ln \le(\frac{\Theta_\Delta(p-x)}{\Theta_{\Delta}(p)} \ri)
\eea
where we have used that $|z(x)|>R$.
\QED

\begin{lemma}
\label{lemmaFay}
Let us pose
\be
\label{Hn}
\scr H_n(\vec r;\bt):= {\rm e}^{Q(\bt)} 
\Theta(\mathbb F) \det \big[\zeta_{j-1}(r_k;\bt) \big]_{j,k=1}^n
\ee
where $\mathbb F := \mathbb F(\bt)$ (in \eqref{defF}).
Then we have 
\bea 
\scr H_n(\vec r;\bt)=K_n {\rm e}^{Q(\bt)}\Theta\le (\sum_{j\leq n } r_j + \mathbb F\ri)  \frac{
\ds \prod_{j<k} \Theta_{\Delta}(r_j-r_k)} 
{\ds \prod_{j\leq n} \Theta_{\Delta}( r_j)^{n-1}{\rm e}^{-\vartheta(r_j;\bt)} \Theta(r_j-\scr D-\K)}
, \label{335}
\eea
where the constant $K_n$ is independent of $\bt$ and it is given by ($\CCC$ defined in \eqref{defCCC})
\be
K_n = {(-\CCC )^{n(n-1)/2}}{\Theta(\scr D+\K)^{n-1}}.
\ee
\end{lemma}
{\bf Proof.}
Both sides  of \eqref{335} are skew--symmetric single--valued functions of the $n$--tuple $(r_1,\dots , r_n)\in \mathcal C^n$, so it suffices to consider them as a function of --say-- $r_n$.  As such, both sides behave near $\infty$ as $z^{n-1}(r_n){\rm e}^{\xi(r_n;\bt)} (1+ \mathcal O(1/z))$ and have poles at $\scr D$ and zeros at $r_1,\dots, r_{n-1}$. There are $g$ other  zeros in generic position. The ratio of the two sides is thus a meromorphic function with $g$ poles (and $g$ zeros) in generic position, but this can only be a constant (by Riemann--Roch's theorem). To compute this constant we proceed by induction on $n$; for $n=1$ the statement is a tautology with $K_1=1$. 
Consider now the induction step; multiply both sides by ${\rm e}^{-\xi(r_n)} z(r_n)^{1-n}$ and send $r_n$ to $\infty$. In the left side the limit is the coefficient in the Laplace expansion of the determinant in front of $\zeta_{n-1}(r_n)$. This is exactly the same determinant in one variable less. We now turn to the right side of \eqref{335}, which we denote by $R_n(r_1,\dots, r_n)$.  We have
\bea
\lim_{r_n\to\infty} \frac { {\rm e}^{-\xi(r_n;\bt)}R_n(r_1,\dots, r_n)}{z(r_n)^{n-1}} =K_n {\rm e}^{Q(\bt)}\Theta\le (\sum_{j\leq n-1} r_j + \mathbb F\ri)  \frac{
\ds \prod_{j<k\leq n-1} \Theta_{\Delta}(r_j-r_k)} 
{\ds \prod_{j\leq n-1} \Theta_{\Delta}( r_j)^{n-2}{\rm e}^{-\vartheta(r_j;\bt)} \Theta(r_j-\scr D-\K)}\times \cr
\times
\frac 1{\Theta (\scr D+\K)}\lim_{r_n\to\infty} \frac 1{z(r_n)^{n-1} \Theta_\Delta(r_n)^{n-1}}
= R_{n-1}(r_1,\dots, r_{n-1}) \frac{ K_{n}}{K_{n-1}\Theta(\scr D+\K)(-\CCC )^{n-1}}
\eea
Therefore we must have  $K_n = {K_{n-1}\Theta(\scr D+\K)}{(-\CCC )^{n-1}}$.
The proof then follows by induction.  
\QED
%We observe that the constant $\CCC $ has also the property that $\omega_\Delta(p) = \frac {\CCC \d z(p)}{z(p)^2}\le( 1 + \mathcal O(z(p)^{-1}) \ri)$ as $p\to\infty\in \mathcal C$ (and $z(p)\to\infty\in \mathbb P^1$).
%\\

%
%
%
%

%%
%\appendix
%\section{Proofs}
%\label{proofs}
%
%
\subsection{ Proof of Proposition \ref{propBaker}}
\label{proofpropBaker}
With the definition \eqref{Hn} the tau--function can be written as 
\be
\label{tau3}
\tau_n(\bt,\bfs) =\frac {{\rm e}^{nA(\bt) + nA(\bfs)}} {n!} \int_{\gamma^n} \prod_j \d\mu (r_j) \scr H_n(\vec r; \bt) \scr H_n(\vec r; \bfs).
\ee
Combining \eqref{Gshift}, \eqref{Qshift+}, Lemma \ref{lemmaV} and the definition of $\mb F$ \eqref{defF} we see that the following holds
\bea
\label{H-}
\scr H_n&( \vec r; \bt - [x]){\rm e}^{\xi(x;\bt)} =K_n
{\rm e}^ {Q(\bt) + \vartheta(x;\bt)} \frac 1{ \Theta_\Delta(x)}\sqrt{\frac{\omega_\Delta(x)}{\d z(x)}} \Theta\le( \sum_j r_j + x +\mb F \ri)  \times \cr 
&\times
\frac{
\ds \prod_\ell \Theta_\Delta(r_\ell-x) \prod_{j<k} \Theta_{\Delta}(r_j-r_k)} 
{\ds \prod_{j} \Theta_{\Delta}( r_j)^{n}{\rm e}^{-\vartheta(r_j;\bt)} \Theta(r_j-\scr D-\K)} 
\eea
where  $\scr H_n$ is defined in \eqref{Hn}.
The expression \eqref{H-} can be rewritten as follows: 
\bea
\label{Hn2}
\scr H_n &( \vec r; \bt - [x]){\rm e}^{\xi(x;\bt)} =\frac {K_n}{K_{n+1}}\sqrt{\frac{\omega_\Delta(x)}{\d z(x)}} \Theta_\Delta(x)^{n-1}\Theta(x-\scr D-\K) \scr H_{n+1}((\vec r,x);\bt)
\eea
We now integrate \eqref{Hn2} with respect to the variables $r_j$'s on $\gamma$ using the formula \eqref{tau3} for the tau functions;
\bea 
\tau(\bt-[x];\bfs) {\rm e}^{\xi(x;\bt)} &=\frac{{\rm e}^{nA(\bt-[x])+nA(\bfs)}}{n!}{\rm e}^{\xi(x;\bt)} 
\int_{\gamma^n} \scr H_n ( \vec r; \bt - [x])\scr H_n(\vec r;\bfs) \prod_{j=1}^n \d\mu(r_j).
\label{tmp110}
\eea
Consider the integral in \eqref{tmp110}: if we use \eqref{Hn2} for $\scr H_n ( \vec r; \bt - [x])$ together with the definition \eqref{Hn} of $\scr H_n$ as a determinant,  the computation to be performed involves now (up to inessential multiplicative constants)
\be
\int_{\gamma^n} \scr H_{n+1} ( (\vec r,x); \bt)\scr H_n(\vec r;\bfs) \prod_{j=1}^n \d\mu(r_j)
= \int_{\gamma^n} \det[\zeta_{a-1}(r_b)]_{a,b=1}^{n+1} \det[\zeta_{a-1}(r_b)]_{a,b=1}^{n} \prod_{j=1}^n \d\mu(r_j)\label{tmp111}
\ee
where, for notational simplicity, in the first determinant we have set $r_{n+1}=x$ (and this variable is not integrated upon). Expanding this determinant along the last row and then using Andreief's identity on each of the coefficients, we obtain $n!P_n(x;\bt,\bfs)$ as in \eqref{defPn}.
Collecting now the factors in \eqref{Hn2}  and this last computation together with \eqref{shiftAn}, the equation  \eqref{tmp110} becomes
\bea
\tau(\bt-[x];\bfs) {\rm e}^{\xi(x;\bt)}= \frac {\star}{\Theta_\Delta(x)} \sqrt{\frac{\omega_\Delta(x)}{\d z(x)}}  \frac{P_n(x;\bt,\bfs)}{z(x)^n}.
\eea
The constant $\star$ can be computed by carefully tracking all the steps or simply by asymptotic analysis near $\infty$; indeed we see from \eqref{defPn} that 
\be
\lim_{x\to \infty} {\rm e}^{-\xi(x;\bt)} \frac{P_n(x;\bt ,\bfs)}{z(x)^n}= \det\bigg[\mu_{ab}(\bt,\bfs)\bigg]_{a,b=0}^{n-1}
\ee
Then also we need that (recall that $\d z(x)$ has a second order pole at $\infty$, while $\omega_\Delta$ is holomorphic, so that the square root of their ratio is a locally defined function with a simple zero)
\be
\lim_{x\to\infty} \frac {-1}{\CCC \Theta_\Delta(x)} \sqrt{\frac{\CCC \omega_\Delta(x)}{\d z(x)}}= 1
\ee
 The conclusion follows from elementary algebra.
\QED
\subsection{ Proof of Proposition \ref{propdualBaker}}
\label{proofpropdualBaker}
Similarly to the proof of Proposition \ref{propBaker}, combining \eqref{Gshift}, \eqref{Qshift-} and Lemma \ref{lemmaV} we have

\bea
\label{H+}
\scr H_n&( \vec r;  \bt + [x]){\rm e}^{-\xi(x; \bt)} ={\rm e}^ {Q( \bt) - \vartheta(x; \bt)} \frac 1{ \Theta_\Delta(x)}\sqrt{\frac{\omega_\Delta(x)}{\d z(x)}} \Theta\le( \sum_j r_j - x + {\mb F}\ri) \times \cr 
&\times \frac{
\ds   \prod_{j<k} \Theta_{\Delta}(r_j-r_k)} 
{\ds \prod_{j} \Theta_{\Delta}( r_j)^{n-2}\Theta_\Delta(r_j-x){\rm e}^{-\vartheta(r_j; \bt)} \Theta(r_j-\scr D-\K)} 
\eea
Now consider the expression
\bea
\phi_n(x;\vec r;\bt)=&{\rm e}^ { - \vartheta(x; \bt)} \Theta\le( \sum_j r_j - x + {\mb F}\ri)\Theta(x-\scr D-\K)\omega_\Delta(x) \times \cr 
&\times \frac{ \Theta_\Delta(x)^{n-2} 
\ds   \prod_{j<k} \Theta_{\Delta}(r_j-r_k)} 
{\ds \prod_{j} \Theta_{\Delta}( r_j)^{n-2}\Theta_\Delta(r_j-x){\rm e}^{-\vartheta(r_j; \bt)} \Theta(r_j-\scr D-\K)} \label{phin}.
\eea
 With respect to $x$, $\phi_n$  is a single--valued one--form in $ \mathcal K_{-\bt}( -\scr D +\sum r_j - (n-2)\infty)$, while, with respect to each of the $r_j$ it is a section of $\scr L_{ \bt}(\scr D+(n-2)\infty)$ and it is skew--symmetric under the action of permutations of the $r_j$ variables. Using the same logic as in the proof of Lemma \ref{lemmaFay} we deduce that, up to a constant, the following holds
\bea
\phi_n(x,\vec r;\bt)\propto \det \le[
\begin{array}{c|cccc}
\Cauchy(x,r_1;\bt) & \zeta_0(r_1;\bt) & \zeta_1(r_1;\bt) & \cdots &\zeta_{n-2} (r_1;\bt)\\
\vdots & \\
\Cauchy(x,r_n;\bt) & \zeta_0(r_n;\bt) & \zeta_1(r_n;\bt) & \cdots &\zeta_{n-2} (r_n;\bt)
\end{array}
\ri]\label{phindet}
\eea
To verify that the behaviour near $x=\infty$ of the right side of \eqref{phindet} is indeed the correct one we use the following local analysis: from the formula for $\zeta_a(r;\bt)$ it follows that 
\be
\zeta_a(r;\bt) = \res{p=\infty}\Cauchy(p,r;\bt ) z(p)^{a}{\rm e}^{\xi(p;\bt)} \ \ \Rightarrow\ \ \
\Cauchy(x,r;\bt) = {\rm e}^{-\xi(x;\bt)}\sum_{\ell=0}^\infty \frac {\d z(x)}{z(x)^{\ell+1}}\zeta_\ell(r;\bt)
\ee
with the series converging for $x,r$ in a neighbourhood of $\infty$ as long as $|z(x)|>|z(r)|$. Inserting this expansion in the first column of the determinant \eqref{phindet} one sees that all the coefficients of $\frac{\d z(x)}{z(x)^a}$ vanish up to $a=n-1$ included and 
\be
\phi_n(x;\vec r;\bt) \propto {\rm e}^{-\xi(x;\bt)} \frac {\d z(x)}{z(x)^n} \det\bigg[\zeta_{a-1}(r_b;\bt) \bigg]_{a,b=1}^{n}\bigg(1 + \mathcal O(z(x)^{-1})\bigg),\ \  x\to\infty.
\label{tmp119}
\ee
 
We now integrate $\phi_n$ as follows
\be
\mathfrak R_n(x;\bt,\bfs):= \frac 1{n!}\int_{\gamma^n}  \phi_n(x;\vec r;\bt) \det\big[\zeta_{a-1}(r_b;\bfs)\big]_{a,b=1}^{n} \prod_j \d\mu(r_j).
\ee
We want to identify $\mathfrak R_n$ with a more transparent formula involving the Cauchy transform of the biorthogonal section $Q_{n-1}$ defined in \eqref{defPn}; to this end we observe that $\mathfrak R_n$ is holomorphic on $\CC\setminus \gamma \cup\{\infty\}$ and has a jump discontinuity across $\gamma$ which we can compute with the aid of Sokhotski--Plemelj formula together with the symmetry under permutation of the dummy variables $r_j$:
\bea
\mathfrak R_n(x;\bt,\bfs)_+-\mathfrak R_n(x;\bt,\bfs)_- = \frac{n 2 \pi i}{n!} \int_{\gamma^{n-1}} \hspace{-10pt}  \det\big[\zeta_{a-1}(r_b;\bt)\big]_{a,b=1}^{n-1} \det\big[\zeta_{a-1}(r_b;\bfs)\big]_{a,b=1}^{n} \prod_{j=1}^{n-1} \d\mu(r_j);\nn \\r_n=x\in \gamma.
\eea
Using the same reasoning that followed \eqref{tmp111} we arrive at the following formula for the integral: 
\be
\mathfrak R_n(x;\bt,\bfs)_+-\mathfrak R_n(x;\bt,\bfs)_- = 2i\pi Q_{n-1}(x;\bt,\bfs)\d\mu(x),\ \ \ x\in \gamma
\ee
Using the fact that $\mathfrak R_n$ vanishes at the non-special divisor $\scr D$ (and the behaviour at $\infty$) we then conclude that  
\be
\mathfrak R_n(x;\bt,\bfs) = \int_{\gamma} \Cauchy(x,r;\bt) Q_{n-1}(r;\bt,\bfs)\d\mu(r).
\ee
From \eqref{tmp119} it follows that near $\infty$ the differential $\mathfrak R_n$ has the expansion
\be
\label{Rnexp}
\mathfrak R_n(x;\bt,\bfs) = n!\frac{{\rm e}^{-\xi(x;\bt)} \d z(x)}{z(x)^{n}} \det\bigg[\mu_{ab}(\bt,\bfs)\bigg]_{a,b=0}^{n-1}\le(1 + \mathcal O\le(\frac1{z(x)}\ri)\ri).
\ee 
Now we can collect this information in the final computation:
\bea
\tau(\bt+[x], \bfs){\rm e}^{-\xi(x;\bt)} =\frac {{\rm e}^{nA( \bt+[x])+ nA(\bfs)} }{n!}\int_{\gamma^n} \scr H_n&( \vec r;  \bt + [x]){\rm e}^{-\xi(x; \bt)} \scr H_n( \vec r; \bfs )  \prod_j\d\mu(r_j).
\eea
Using \eqref{shiftAn}, \eqref{H+} and the resulting expression of $\mathfrak R_n$ we obtain, up to a multiplicative constant to be determined
\be
\tau(\bt+[x], \bfs){\rm e}^{-\xi(x;\bt)} \propto z(x)^{n}  \mathfrak R_n(x;\bt,\bfs) { \Theta_\Delta(x)}\sqrt{\frac{\omega_\Delta(x)}{\d z(x)}} 
\ee
The proportionality constant can be evaluated by asymptotic expansion at infinity using \eqref{Rnexp} and \eqref{omeganorm}.
%\be
%\frac { \Theta_\Delta(x)}{-\CCC }\sqrt{\frac{\omega_\Delta(x)}{\CCC \d z(x)}}  \sim  \frac {1}{z(x)^2}
%\ee
\QED

\end{document}